\documentclass[a4paper,english,aps,preprint,superscriptaddress]{article}
\usepackage{epsfig}
\usepackage[]{fontenc}
\usepackage[latin9]{inputenc}
\usepackage{amsmath}
\usepackage{esint}
\usepackage{bm}
\usepackage{rotating}
\usepackage{slashed}
\usepackage{multirow}
\usepackage{graphicx}  
\usepackage{color}     
\usepackage{color}
\usepackage{lscape}
\usepackage{mathrsfs}
\usepackage{pdflscape}
\date{\today}
 \normalsize

\newcommand{\bmat}{\left(\begin{array}}
\newcommand{\emat}{\end{array}\right)}
\newcommand{\be}{\begin{equation}}
\newcommand{\ee}{\end{equation}}
\newcommand{\bea}{\begin{eqnarray}}
\newcommand{\eea}{\end{eqnarray}}

\def\gtwid{\mathrel{\raise.3ex\hbox{$>$\kern-.75em\lower1ex\hbox{$\sim$}}}}
\def\ltwid{\mathrel{\raise.3ex\hbox{$<$\kern-.75em\lower1ex\hbox{$\sim$}}}}
\def\gev{{\rm \, Ge\kern-0.125em V}}
\def\tev{{\rm \, Te\kern-0.125em V}}

\begin{document}

\title{Neutrino Mass Matrices with one texture equality and one vanishing neutrino mass}

\author
{ \it \bf   Jinzhong Han$^1$, Ruihong Wang$^2$, Weijian Wang$^3$\thanks{wjnwang96@aliyun.com} , Xing-Ning Wei$^3$
\\ \small$^1$ School of Physics and Telecommunications Engineering, Zhoukou Normal University, Henan, 466001, China \\
\small $^2$College of Information Science $\&$Technology, Hebei Agricultural University, Baoding,China\\
\small$^3$ Department of Physics, North China Electric Power University, Baoding 071003, P. R. China \\}

\maketitle

\begin{abstract}
In the light of latest data of neutrino oscillation experiments, we carry out a systematic investigation on the texture structures of Majorana neutrino mass matrix $M_{\nu}$, which contain one vanishing neutrino mass and an equality between two matrix elements.  Among 15 logically possible patterns, it is found that for norm order ($m_{3}>m_{2}>m_{1}=0$) of neutrino masses only five of them are compatible with recent experimental data at the $3\sigma$ level, while for inverted order ($m_{2}>m_{1}>m_{3}=0$) ten patterns is phenomenologically allowed. In the numerical analysis, we perform a scan over the parameter space of all viable patterns to get a large sample of scattering points. We present the implications of each allowed pattern for three mixing angles $(\theta_{12},\theta_{2},\theta_{3})$, leptonic CP violation and neutrinoless double-beta decay, predicting strong correlations between oscillation parameters. The theoretical realization of a concrete example is discussed in the framework of Froggatt-Nielsen mechanism.

\vspace{1em}

\end{abstract}
\maketitle

\section{Introduction}

In spite of the evidences for massive neutrinos and the admixtures of the flavor states \cite{neu1}, the origin of the lepton flavor structure remains an open question.  A popular approach to
understand the leptonic mixing structure is reducing the number of free parameters by adding Abelian or non-Abelian flavor symmetries, leading to specific texture structures of neutrino mass matrix ($M_{\nu}$).
Some common models include
texture-zeros\cite{zero,zn,zero1,GCB,z,z1}, hybrid textures\cite{hybrid,
hybrid2}, zero trace\cite{sum},
vanishing minors\cite{minor1,minor,minor2}, two traceless
submatrices\cite{tra}, equal elements or cofactors\cite{co}, hybrid
$M_{\nu}^{-1}$ textures\cite{hyco}, partial $\mu-\tau$ symmetry\cite{Lashin:2013xha}.

One can also reduce the number of free parameters of $M_{\nu}$ by assuming one of neutrinos to be massless.  It is well known that in type-I seesaw mechanism,  a vanishing neutrino mass is easily realized by assuming two families of heavy right-handed
neutrinos. Similar scenario also appears in radiative seesaw model\cite{Wang:2015saa,Wang:2017mcy}. In the framework of Affleck-Dine mechanism\cite{Affleck:1984fy,Murayama:1993em,Dine:1995kz}, an extremely small neutrino mass($ \approx 10^{-10}$eV) is required to successfully produce the leptogenesis\cite{Asaka:2000nb,Fujii:2001sn}.  Hence it is imperative to ask if the lepton mass matrix with a specific texture structure and a vanishing mass eigenvalue can survive under the current neutrino oscillation data. Several attempts have been made in this direction. In Refs.\cite{Antusch:2011ic,Rodejohann:2012jz,Gautam:2015kya}, particular attentions has been paid to the neutrino mass matrix structures with one texture-zero and one massless eigenstate.

In the flavor basis where the charged lepton mass matrix is diagonal, we investigate a specific class of neutrino mass matrices which contains an vanishing neutrino mass and an texture equality between two independent entries. The texture equality was proposed in many literatures. Bear in mind the fact that an equality between "12" and "13" element of $M_{\nu}$ has appeared in the the tri-bimaximal form of $M_{\nu}$
\begin{equation}
M_{\nu}=\left(\begin{array}{ccc}
a&b&b\\
b&a-c&b+c\\
b&b+c&a-c
\end{array}\right)
\label{1}\end{equation}
which can be diagonalized by tri-bimaximal form of PMNS matrix
\begin{equation}
U_{\text{PMNS}}=\left(\begin{array}{ccc}
\sqrt{\frac{2}{3}}&\sqrt{\frac{1}{3}}&0\\
-\sqrt{\frac{1}{6}}&\sqrt{\frac{1}{3}}&\sqrt{\frac{1}{2}}\\
\sqrt{\frac{1}{6}}&\sqrt{\frac{1}{3}}&-\sqrt{\frac{1}{2}}
\end{array}\right)
\label{tri}\end{equation}
and leads to the three neutrino masses eigenvalues $m_{1}=a-b, m_{2}=a+2b, m_{3}=a-b-2c$. 
If $| a|\simeq |b| \ll |c|$, we obtain a normal hierarchy for neutrino masses with $m_{1}\simeq 0$. Clearly, Eq.\eqref{tri} is not compatible with the observation of no-zero $\theta_{13}$ angle. Hence the texture structure with an equality between  "12" and "13" elements of $M_{\nu}$ can be considered as a perturbed generalization of Eq.\eqref{1},  where the number of free parameters is added but the texture equality still holds.

The main motivation of this work is three-fold:
\begin{itemize}
\item From the phenomenological viewpoint,  either a vanishing neutrino mass or an equality between two nozero matrix elements imposes one constraint condition on $M_{\nu}$ and reduces the number of free degrees by two. Thus texture equalities are as predictive as the texture zeros.
\item On the experimental side, the absolute scale of lightest neutrino mass is still unknown.  Within the $\Lambda$CDM framework , an upper bound on the sum of neutrino mass $\sum m_{i}<0.23$ eV at $95\%$ confidence level has been reported by Plank Collaboration\cite{Planck} . Recently,  combined with the Planck, TT,TE,EE+lowP+BAO+JLA+H0+Lensing data, a much tighter bound $\sum m_{i}<0.105$, which is almost access to the lower limit of $\sum m_{i}$ for inverted order spectrum of neutrino masses, is obtained in holographic dark energy scenario\cite{Wang:2016tsz}. Since we have stood the  on the verge to distinguish the mass spectrum (NO or IO) of three neutrinos through cosmological observations, the phenomenology of some specific texture structures with a vanishing neutrino mass deserves a detailed survey. 
\item It is generally believed that the observed neutrino mixing pattern indicates some
underlying discrete flavor symmetries.  The flavor symmetry realization for all possible patterns is beyond the scope of this work.  In the framework of the Froggatt-Nielsen (FN) mechanism\cite{FN}, we consider an explicit model based on $S_{\mu-\tau}\times Z_{8}$ symmetry suggested by Ref.\cite{Lashin:2013xha} to realize one of the viable texture equality.  Then we adopt the methodology in Ref.\cite{Fujii:2001zr} by assuming the broken FN symmetry a discrete $Z_{n}$. We will see that a ultalight neutrino mass naturally arises as a good approximation to vanishing neutrino masses. We expect that a phenomenological analysis may help us reveal the underlying flavor structure of lepton mixing.
\end{itemize}

The rest of paper is organized as follows: In Sec.II, We describe some
useful notations and the framework used to obtain the constraint
equations. In Sec. III, the details of numerical
analysis are presented. In Sec. IV, we discuss the theoretical realization for a concrete example. The summary is given in Sec. V.
\begin{table}[!htbp]
\caption{Fifteen possible texture structures with equality between two nozero elements where "$\bigtriangleup$" denotes the nozero and equal elements, while "$\times$" stands for the abitrary and nozero ones.}\label{o1}
\begin{tabular}{c c c c c c }
\hline\hline \quad $P1$ \quad &$P2$\quad &\quad $P3$
\quad
 &\quad  $P4$ \quad  & \quad  $P5$ \quad \\
$\left(\begin{array}{ccc}
\bigtriangleup&\bigtriangleup&\times\\
\bigtriangleup&\times&\times\\
\times&\times&\times
\end{array}\right)$ & $\left(\begin{array}{ccc}
\bigtriangleup&\times&\bigtriangleup\\
\times&\times&\times\\
\bigtriangleup&\times&\times
\end{array}\right)
$& $\left(\begin{array}{ccc}
\times&\bigtriangleup&\times\\
\bigtriangleup&\bigtriangleup&\times\\
\times&\times&\times
\end{array}\right)$& $\left(\begin{array}{ccc}
\times&\times&\times\\
\times&\bigtriangleup&\bigtriangleup\\
\times&\bigtriangleup&\times
\end{array}\right)$& $\left(\begin{array}{ccc}
\times&\times&\bigtriangleup\\
\times&\times&\times\\
\bigtriangleup&\times&\bigtriangleup
\end{array}\right)$ \\
 \quad $P6$ \quad &$P7$\quad &\quad $P8$
\quad
 &\quad  $P9$ \quad  & \quad  $P10$  \\
$\left(\begin{array}{ccc}
\times&\times&\times\\
\times&\times&\bigtriangleup\\
\times&\bigtriangleup&\bigtriangleup
\end{array}\right)$& $\left(\begin{array}{ccc}
\bigtriangleup&\times&\times\\
\times&\times&\bigtriangleup\\
\times&\bigtriangleup&\times
\end{array}\right)$& $\left(\begin{array}{ccc}
\times&\times&\bigtriangleup\\
\times&\bigtriangleup&\times\\
\bigtriangleup&\times&\times
\end{array}\right)$& $\left(\begin{array}{ccc}
\times&\bigtriangleup&\times\\
\bigtriangleup&\times&\times\\
\times&\times&\bigtriangleup
\end{array}\right)$& $\left(\begin{array}{ccc}
\bigtriangleup&\times&\times\\
\times&\bigtriangleup&\times\\
\times&\times&\times
\end{array}\right)$\\
\quad $P11$ \quad &$P12$\quad &\quad $P13$
\quad
 &\quad  $P14$ \quad  & \quad  $P15$ \quad   \\
$\left(\begin{array}{ccc}
\bigtriangleup&\times&\times\\
\times&\times&\times\\
\times&\times&\bigtriangleup
\end{array}\right)$ & $\left(\begin{array}{ccc}
\times&\times&\times\\
\times&\bigtriangleup&\times\\
\times&\times&\bigtriangleup
\end{array}\right)$& $\left(\begin{array}{ccc}
\times&\bigtriangleup&\bigtriangleup\\
\bigtriangleup&\times&\times\\
\bigtriangleup&\times&\times
\end{array}\right)$& $\left(\begin{array}{ccc}
\times&\bigtriangleup&\times\\
\bigtriangleup&\times&\bigtriangleup\\
\times&\bigtriangleup&\times
\end{array}\right)$& $\left(\begin{array}{ccc}
\times&\times&\bigtriangleup\\
\times&\times&\bigtriangleup\\
\bigtriangleup&\bigtriangleup&\times
\end{array}\right)$ \\
\hline\hline
\end{tabular}
\end{table}

\section{Formalism and important relations}
Assuming neutrinos the Majorana particles as proposed in various seesaw models\cite{seesaw}, the neutrino mass matrix $M_{\nu}$ is a symmetric and generally complex matrix with six independent entries.  We arrive at $C_{6}^{2}=15$ logically possible patterns to place an equality between two matrix elements, which are shown in Table.I.

In following analysis, we conider the Pontecorvo-Maki-Nakagawa-Sakata matrix
$U_{\text{PMNS}}$\cite{PMNS} parameterized as
\begin{equation} U_{\text{PMNS}}=UP_{\nu}=\left(\begin{array}{ccc}
c_{12}c_{13}&s_{12}c_{13}&s_{13}\\
-c_{12}s_{23}s_{13}-s_{12}c_{23}e^{-i\delta}&-s_{12}s_{23}s_{13}+c_{12}c_{23}e^{-i\delta}&c_{13}s_{23}\\
-c_{12}c_{23}s_{13}+s_{12}s_{23}e^{-i\delta}&-c_{23}s_{12}s_{13}-c_{12}s_{23}e^{i\delta}&c_{13}c_{23}
\end{array}\right)\left(\begin{array}{ccc}
e^{2i\rho}&0&0\\
0&e^{2i\rho}&0\\
0&0&1
\end{array}\right)
\label{3}\end{equation} where the abbreviations
$s_{ij}=\sin\theta_{ij}$ and $c_{ij}=\cos \theta_{ij}$ are used. The
($\rho$,$\sigma$) in $P_{\nu}$ denotes two Majorana CP-violating
phases and $\delta$ stands for Dirac CP-violating phase.

In the flavor basis where $M_{l}$ is diagonal, the Majorana neutrino mass $M_{\nu}$ is related to the diagonal mass matrix $M^{D}=$diag$(m_{1},m_{2},m_{3})$ though the unitary transformation
\begin{equation}
M_{\nu}=U_{\text{PMNS}}M^{D}U_{\text{PMNS}}^{T}
\end{equation}
and can be re-expressed as
\begin{equation}
M_{\nu}=U\left(\begin{array}{ccc}
\lambda_{1}&0&0\\
0&\lambda_{2}&0\\
0&0&\lambda_{3}
\end{array}\right)U^{T}
\end{equation}
where $\lambda_{1}=m_{1}e^{2i\rho},\lambda_{2}=m_{2}e^{2i\sigma},\lambda_{3}=m_{3}$. In this parametrization, the mass matrix elements are given by
\begin{equation}\begin{split}
(M_{\nu})_{11}=&m_{1}c_{12}^2 c_{13}^2 e^{2i\rho}+m_{2}s_{12}^2c_{13}^2 e^{2i\sigma}+m_{3}s_{13}^2\\
(M_{\nu})_{12}=&m_{1}(-c_{12}^2 s_{23}c_{13}s_{13} e^{2i\rho}-c_{12}s_{12}c_{23}c_{13} e^{i(2\rho-\delta)})          
                         +m_{2}(-s_{12}^2 s_{23}c_{13}s_{13} e^{2i\sigma}+c_{12} s_{12}c_{23}c_{13} e^{i(2\sigma-\delta)})\\
                         &+m_{3}s_{23}c_{13}s_{13}\\
(M_{\nu})_{13}=&m_{1}(-c_{12}^2 c_{23}c_{13}s_{13} e^{2i\rho}+c_{12}s_{12}s_{23}c_{13} e^{i(2\rho-\delta)})  
                          +m_{2}(-s_{12}^2 c_{23}c_{13}s_{13} e^{2i\sigma}+c_{12} s_{12}s_{23}c_{13} e^{i(2\sigma-\delta)}) \\
                          &+m_{3}c_{23}c_{13}s_{13} \\
(M_{\nu})_{22}=&m_{1}(c_{12} s_{23}s_{13} e^{2i\rho}+c_{23}s_{12}e^{i(\rho-\delta)})^2+m_{2}(s_{12} s_{23}s_{13}e^{2i\sigma}-c_{23}c_{12}e^{i(\sigma-\delta)})^2+m_{3}c_{13}^2s_{23}^2 \\
 (M_{\nu})_{23}=&m_{1}(c_{12}^2 c_{23} s_{23}s_{13}^2 e^{2i\rho}+c_{12}s_{12}s_{13}(c_{23}^2-s_{23}^2)e^{i(2\rho-\delta)}-c_{23}s_{23}s_{12}^2e^{2i(\rho-\delta)})\\
                           &+m_{2}(s_{12}^2 c_{23} s_{23}s_{13}^2 e^{2i\sigma}+c_{12}s_{12}s_{13}(s_{23}^2-c_{23}^2)e^{i(2\sigma-\delta)}-c_{23}s_{23}c_{12}^2e^{2i(\sigma-\delta)})+m_{3}c_{23}s_{23}c_{13}^2\\   
(M_{\nu})_{33}=&m_{1}(c_{12} c_{23}s_{13} e^{2i\rho}- s_{23}s_{12}e^{i(\rho-\delta)})^2 +m_{2}(s_{12} c_{23}s_{13}e^{2i\sigma}+s_{23}c_{12}e^{i(\sigma-\delta)})^2+m_{3}c_{13}^2c_{23}^2                                                                        
\end{split}\end{equation}

\begin{table}[!htbp]
\caption{The latest global-fit results of neutrino oscillation parameters given by Ref.\cite{Capozzi:2017ipn} , with $\delta m^{2}=m_{2}^{2}-m_{1}^{2}$, $\Delta m^{2}=|m_{3}^{2}-\frac{1}{2}(m_{2}^{2}+m_{1}^{2}|)$.  }\label{o2}
\begin{tabular}{c c c c c c }
\hline\hline
Parameter & Order&  $1\sigma$ range & $2\sigma$ range & $3\sigma$ range \\
\hline
$\delta m^2/10^{-5}~\mathrm{eV}^2 $ & NO, IO  & 7.21 - 7.54 & 7.07 - 7.73 & 6.93 - 7.96 \\
\hline
$\sin^2 \theta_{12}/10^{-1}$ & NO, IO  & 2.81 - 3.14 & 2.65 - 3.34 & 2.50 - 3.54 \\
\hline
$|\Delta m^2|/10^{-3}~\mathrm{eV}^2 $ & NO  & 2.495 - 2.567 & 2.454 - 2.606 & 2.411 - 2.646 \\
                                      & IO  & 2.473 - 2.539 & 2.430 - 2.582 & 2.390 - 2.624 \\
\hline
$\sin^2 \theta_{13}/10^{-2}$ & NO  & 2.08 - 2.22 & 1.99 - 2.31 & 1.90 - 2.40 \\
                             & IO  & 2.07 - 2.24 & 1.98 - 2.33 & 1.90 - 2.42 \\
\hline
$\sin^2 \theta_{23}/10^{-1}$ & NO  & 4.10 - 4.46 & 3.95 - 4.70 & 3.81 - 6.15 \\
                             & IO  & 4.17 - 4.48 $\oplus$ 5.67 - 6.05 & 3.99 - 4.83 $\oplus$ 5.33 - 6.21 &  3.84 - 6.36 \\
\hline
$\delta/\pi$ & NO  & 1.18 - 1.61 & 1.00 - 1.90  &  0 - 0.17 $\oplus$ 0.76 - 2   \\
             & IO  & 1.12 - 1.62 & 0.92 - 1.88  &  0 - 0.15 $\oplus$ 0.69 - 2  \\
\hline\hline
\end{tabular}
\end{table}

In terms of the texture equality condition [e.g $(M_{\nu})_{ab}=M_{\nu})_{cd}$], we obtain the constraint condition equation
\begin{equation}
 \sum_{n=1}^3(U_{ai}U_{bi}-U_{ci}U_{di})\lambda_{i}=0
\label{equal}\end{equation}
which, if the lightest neutrino is massless, leads to
\begin{equation}\begin{split}
&\xi\equiv\frac{m_{2}}{m_{3}}=\left|\frac{U_{a3}U_{b3}-U_{c3}U_{d3}}{U_{a2}U_{b2}-U_{c2}U_{d2}}\right|\\
&\sigma=-\frac{1}{2}arg\left(\frac{U_{a3}U_{b3}-U_{c3}U_{d3}}{U_{a2}U_{b2}-U_{c2}U_{d2}}\right)
\end{split}\label{c1}\end{equation}
for normal order (NO) spectrum. The three neutrino masses are given by
\begin{equation}
m_{1}=0,\quad m_{2}=\sqrt{\delta m^{2}},\quad m_{3}=\frac{m_{2}}{\xi}
\label{m1}\end{equation}
It is clear that, the Majorana CP-violating phase $\rho$ becomes unphysical, since $m_{1}$ is zero, and can be dropped out in numerical calculation.  

For invert order (IO) spectrum, we obtain
\begin{equation}\begin{split}
&\zeta\equiv\frac{m_{2}}{m_{1}}=\left|\frac{U_{a1}U_{b1}-U_{c1}U_{d1}}{U_{a2}U_{b2}-U_{c2}U_{d2}}\right|\\
&\rho-\sigma=-\frac{1}{2}arg\left(\frac{U_{a1}U_{b1}-U_{c1}U_{d1}}{U_{a2}U_{b2}-U_{c2}U_{d2}}\right)
\end{split}\label{c2}\end{equation}
and three neutrino masses given by
\begin{equation}
m_{3}=0,\quad m_{1}=\Delta m^{2}-\frac{\delta m^{2}}{2},\quad m_{2}=\zeta m_{2}
\label{m2}\end{equation}
Note that only the difference $(\rho-\sigma)$ is physical in this case. Hence the neutrino mass ratios and Majorana CP-violating phases are fully determined in terms of the mixing angles $(\theta_{12},\theta_{23},\theta_{13})$ and the Dirac-CP violating phase $\delta$.  The three mixing angles $(\theta_{12},\theta_{23},\theta_{13})$ as well as two independent neutrino mass-squared differences $\delta m^{2}=m_{2}^{2}-m_{1}^{2}$, $\Delta m^{2}=|m_{3}^{2}-\frac{1}{2}(m_{2}^{2}+m_{1}^{2})|$  are precisely measured by many neutrino oscillation experiments. We summarize the latest global-fit results\cite{Capozzi:2017ipn} of neutrino oscillation parameters in Table.II. One can further define the ratio of neutrino mass-squared difference as
\begin{equation}
R_{\nu}=\frac{\delta m^{2}}{\Delta m^{2}}
\end{equation}
which, using Eq.\eqref{c1} and \eqref{c2}, can be expressed as
\begin{equation}
R_{\nu}=\left( \left|\frac{U_{a2}U_{b2}-U_{c2}U_{d2}}{U_{a3}U_{b3}-U_{c3}U_{d3}}\right|^2-1 \right)^{-1}
\label{r1}\end{equation}
in case of normal order and
\begin{equation}
R_{\nu}=1-\left|\frac{U_{a2}U_{b2}-U_{c2}U_{d2}}{U_{a1}U_{b1}-U_{c1}U_{d1}}\right|^2
\label{r2}\end{equation}
in case of inverted order.

The Dirac CP-violation in
neutrino oscillation experiments can be described by the Jarlskog
rephasing invariant quantity
\begin{equation}
J_{CP}=s_{12}s_{23}s_{13}c_{12}c_{23}c_{13}^{2}sin\delta
\end{equation}
On the other hand, the Majorana CP violation can be established if any signal of neutrinoless double beta
($0\nu\beta\beta$) decay is observed. The rate of $0\nu\beta\beta$
decay is determined by the effective Majorana neutrino mass
$m_{ee}$
\begin{equation}
m_{ee}=\left|m_{1}c_{12}^{2}c_{13}^{2}e^{2i\rho}+m_{2}s_{12}^{2}c_{13}^{2}e^{2i\sigma}+m_{3}s_{13}^{2}\right|
\label{mee}\end{equation}
The next generation
$0\nu\beta\beta$ experiments, with the aimed sensitivity of $ m_{ee}$
being up to 0.01 eV\cite{NDD}, will open the window to weigh both neutrino masses and lepton number violation. Besides the $0\nu\beta\beta$ experiments, more severe
constraints on the absolute scale of neutrino masses is set from cosmology observation.

\section{Numerical analysis}
\subsection{A benchmark point for P1 pattern }
First of all,  we point out that the requirement of one vanishing neutrino mass is equivalent to the zero determinant of neutrino mass matrix Det$(M_{\nu})=0$ because
\begin{equation}\begin{split}
\text{Det} (M_{\nu})&=\text{Det}(U_{\text{PMNS}}M^{D}U_{\text{PMNS}}^{T})=\text{Det}(U_{\text{PMNS}}^{T}U_{\text{PMNS}}M^{D})\\
&=\text{Det}(U_{\text{PMNS}}^{T})\text{Det}(U_{\text{PMNS}})\text{Det}(M^{D})=0
\end{split}\end{equation}
In Ref.\cite{Chauhan:2006uf}, the neutrino mass matrices with one texture equality and Det$(M_{\nu})=0$ has been discussed. The authors claimed that such texture structures are not allowed for $\theta_{13}\neq 0$. However, our analysis demonstrates a different result. Here we take the P1 pattern as an illustration.

For NO spectrum of neutrino masses, we set the benchmark points lying in $3\sigma$ range of experimental data as
\begin{equation}\begin{split}
(\theta_{12},\theta_{23},\theta_{13})=(32.2386^{\circ},40.9554^{\circ},8.4901^\circ)\\
(\delta,\sigma)=(83.7040^{\circ},-25.6372^{\circ})\\
(\delta m^{2},\Delta m^{2})=(7.4156\times 10^{-5}\text{eV}^2,2.500\times 10^{-3}\text{eV}^2)
\end{split}\end{equation}
then the corresponding neutrino mass matrix is 
\begin{equation}
M_{\nu}=\left(\begin{array}{ccc}
0.0026 - 0.0019i&0.0026 - 0.0019i&0.0071 + 0.0020i\\
0.0026 - 0.0019i&0.0187 + 0.0026i&0.0267 - 0.0019i\\
0.0071 + 0.0020i&0.0267 - 0.0019i& 0.0255 + 0.0012i
\end{array}\right)
\end{equation}
with three neutrino masses given by
\begin{equation}
m_{1}=0,\quad m_{2}=0.0086,\quad m_{3}=0.0500
\end{equation}
For IO spectrum of neutrino masses, if benchmark points are taken as following
\begin{equation}\begin{split}
(\theta_{12},\theta_{23},\theta_{13})=(32.8056^{\circ},40.3032^{\circ},8.3978^\circ)\\
(\delta,\rho,\sigma)=(129.1235^\circ,99.0824^{\circ},162.6441^\circ)\\
(\delta m^{2},\Delta m^{2})=(7.1394\times 10^{-5}\text{eV}^{2},2.500\times 10^{-3}\text{eV}^{2})
\end{split}\end{equation}
we can obtain the neutrino mass matrix 
\begin{equation}
M_{\nu}=\left(\begin{array}{ccc}
-0.0209 - 0.0191i&-0.0209 - 0.0191i&0.0218 + 0.0199i\\
-0.0209 - 0.0191i&0.0166 + 0.0156i&-0.0100 - 0.0096i\\
0.0218 + 0.0199i&-0.0100 - 0.0096i& 0.0043 + 0.0043i
\end{array}\right)
\end{equation}
with three  corresponding neutrino masses
\begin{equation}
m_{1}=0.0501,\quad m_{2}=0.0508,\quad m_{3}=0
\end{equation}
It is clearly that, by direct calculation, the P1 pattern of neutrino mass matrix with $M_{\nu11}=M_{\nu12}$ and a vanishing neutrino mass ($m_{1}$ or $m_{3}=0$) is phenomenologically allowed if appropriate oscillation parameters are chosen.

\begin{landscape}
\begin{table}[h]
    \caption{The various predictions for $P1-P15$ patterns. The symbol $\times$ denotes that the corresponding pattern is not phenomenologically allowed under current experimental data.}
{ \begin{tabular}{c c c c c c }
\hline\hline
Textures & Specturm&  $\theta_{23}$  & $\delta$ & Majorana phases &$m_{ee}$(eV) \\
\hline
$P1$ & NO & $38.22^\circ$ - $52.31^\circ$ & $78.8^\circ$ - $300.2^\circ$ & ($-0.507^\circ$) - $0.502^\circ$ & 0.0028 - 0.0040\\
     & IO & $38.36^\circ$ - $52.84^\circ$ & $106.9^\circ$ - $245.1^\circ$ & ($-89.6^\circ$) - ($-59.1^\circ$) $\oplus$ $58.0^\circ$ - $89.4^\circ$ & 0.0226 - 0.0291\\
$P2$ & NO & $37.85^\circ$ - $52.84^\circ$ & $0^\circ$ - $113.7^\circ$ $\oplus$ $252.6^\circ$ - $360^\circ$ & ($-0.513^\circ$) - ($-0.015^\circ$) $\oplus$ $0.002^\circ$ - $0.503^\circ$ & 0.0028 - 0.0041\\
     & IO & $38.41^\circ$ - $52.77^\circ$ & $0^\circ$ - $68.3^\circ$ $\oplus$ $300.1^\circ$ - $360^\circ$ & ($-91.3^\circ$) - ($-53.2^\circ$) $\oplus$ $57.6^\circ$ - $93.3^\circ$ & 0.0233 - 0.0297\\
\hline
$P3$ & NO & $\times$ & $\times$ & $\times$ & $\times$\\
     & IO & $38.5^\circ$ - $53.7^\circ$ & $71.8^\circ$ - $93.3^\circ$ $\oplus$ $271.4^\circ$ - $291.0^\circ$ & ($-33.6^\circ$) - ($-18.7^\circ$) $\oplus$ $18.0^\circ$ - $35.8^\circ$ & 0.0417 - 0.0475\\
$P5$ & NO & $\times$ & $\times$ & $\times$ & $\times$\\
     & IO & $38.2^\circ$ - $53.5^\circ$ & $97.6^\circ$ - $104.7^\circ$ $\oplus$ $254.2^\circ$ - $268.8^\circ$ & ($-37.2^\circ$) - ($-19.1^\circ$) $\oplus$ $18.3^\circ$ - $36.4^\circ$ & 0.0416 - 0.0472\\
\hline
$P4$ & NO & $49.7^\circ$ - $52.2^\circ$ & $0^\circ$ - $360^\circ$ & ($-1.534^\circ$) - $1.567^\circ$ & 0.0012 - 0.0040\\
     & IO & $\times$ & $\times$ & $\times$ &$\times$\\
$P6$ & NO & $38.4^\circ$ - $40.2^\circ$ & $0^\circ$ - $360^\circ$ & ($-1.566^\circ$) - $1.565^\circ$ & 0.0011 - 0.0040\\
     & IO & $\times$ & $\times$ & $\times$ &$\times$\\
\hline
$P7$ & NO & $\times$ & $\times$ & $\times$ &$\times$\\
     & IO & $\times$ & $\times$ & $\times$ &$\times$\\
\hline
$P8$ & NO & $\times$ & $\times$ & $\times$ &$\times$\\
     & IO & $37.3^\circ$ - $52.9^\circ$ & $99.6^\circ$ - $124.7^\circ$ $\oplus$ $227.6^\circ$ - $265.3^\circ$ & ($-58.2^\circ$) - ($-23.0^\circ$) $\oplus$ $20.2^\circ$ - $57.6^\circ$ & 0.0335 - 0.0467\\
$P9$ & NO & $\times$ & $\times$ & $\times$ &$\times$\\
     & IO & $37.1^\circ$ - $53.4^\circ$ & $68.1^\circ$ - $94.2^\circ$ $\oplus$ $277.0^\circ$ - $308.1^\circ$ & ($-60.0^\circ$) - ($-21.8^\circ$) $\oplus$ $23.1^\circ$ - $59.7^\circ$ & 0.0321 - 0.0459\\
\hline
$P10$ & NO & $\times$ & $\times$ & $\times$ &$\times$\\
      & IO & $\times$ & $\times$ & $\times$ &$\times$\\
$P11$ & NO & $\times$ & $\times$ & $\times$ &$\times$\\
      & IO & $\times$ & $\times$ & $\times$ &$\times$\\
\hline
$P12$ & NO & $44.3^\circ$ - $44.5^\circ$ $\oplus$ $45.6^\circ$ - $52.8^\circ$ & $0^\circ$ - $360^\circ$ & ($-1.555^\circ$) - $1.546^\circ$ & 0.0011 - 0.0041\\
      & IO & $38.1^\circ$ - $52.7^\circ$ & $0^\circ$ - $90.1^\circ$ $\oplus$ $99.6^\circ$ - $269.9^\circ$ $\oplus$ $274.8^\circ$ - $360^\circ$ & ($-52.5^\circ$) - $53.0^\circ$ & 0.0340 - 0.0500\\
\hline
$P13$ & NO & $\times$ & $\times$ & $\times$ &$\times$\\
      & IO & $37.3^\circ$ - $43.5^\circ$ $\oplus$ $47.8^\circ$ - $53.4^\circ$ & $0^\circ$ - $83.6^\circ$ $\oplus$ $103.7^\circ$ - $254.9^\circ$ $\oplus$ $289.3^\circ$ - $360^\circ$ & ($-1.26^\circ$) - $1.26^\circ$ & 0.0476 - 0.0501    \\
\hline
$P14$ & NO & $\times$ & $\times$ & $\times$ &$\times$\\
      & IO & $38.4^\circ$ - $53.7^\circ$ & $103.5^\circ$ - $128.2^\circ$ $\oplus$ $226.3^\circ$ - $260.1^\circ$ & ($-58.3^\circ$) - ($-39.0^\circ$) $\oplus$ $37.9^\circ$ - $59.3^\circ$ & 0.0318 - 0.0404\\
$P15$ & NO & $\times$ & $\times$ & $\times$ &$\times$\\
      & IO & $38.3^\circ$ - $53.6^\circ$ & $58.2^\circ$ - $77.8^\circ$ $\oplus$ $288.0^\circ$ - $309.5^\circ$ & ($-58.1^\circ$) - ($-37.9^\circ$) $\oplus$ $36.2^\circ$ - $57.9^\circ$ & 0.0327 - 0.0409\\
\hline\hline
\end{tabular}
 }
 \label{modelc33}
  \end{table}
\end{landscape}

\subsection{Numerical results and discussion}
We have performed a numerical analysis of all fifteen texture structures shown in Table. I.
For each pattern of $M_{\nu}$, the a set of random number inputs are generated for the
three mixing angles $(\theta_{12}, \theta_{23}, \theta_{13})$ and the neutrino mass square
differences ($\delta m^{2}, \Delta m^{2}$) in their $3\sigma$ range (Table. II). Instead, we generate a random input of Dirac CP-violating phase $\delta$ in the range of $[0, 2\pi)$.
From Eqs.\eqref{r1} and \eqref{r2}, $R_{\nu}$ is determined by both ($\theta_{12},\theta_{23},\theta_{13},\delta$).
This requires the input
scattering point of $(\theta_{12},\theta_{23},\theta_{13},\delta)$
is empirically acceptable only when $R_{\nu}$ falls inside the
$3\sigma$ range $[\delta m^{2}_{min}/\Delta m^{2}_{max},\delta
m^{2}_{max}/\Delta m^{2}_{min}]$. From
Eq.\eqref{m1} and \eqref{m2}, we further get the three absolute
scale of neutrino masses $m_{1,2,3}$.  Two Majorana CP-violating
$\alpha$ and $\beta$ can be constrained by Eq. \eqref{r1}, \eqref{r2} and the allowed range of $m_{ee}$ are subsequently obtained. In the numerical analysis, we require the upper bound on the sum of neutrino masses $\sum m_{i}<0.23$ eV being satisfied.

Before proceeding, one notes that the there exists a so-called $\mu-\tau$
permutation transformation that can relate one texture pattern to another though
\begin{equation}
\widetilde{M}_{\nu}=P_{23}M_{\nu}P_{23}
\end{equation}
where
\begin{equation}
P_{23}=\left(\begin{array}{ccc}
1&0&0\\
0&0&1\\
0&1&0
\end{array}\right)
\end{equation}
and the oscillation parameters between
$M_{\nu}$ and $\widetilde{M}_{\nu}$ given by
\begin{equation}
\widetilde{\theta}_{12}=\theta_{12},\quad
\widetilde{\theta}_{13}=\theta_{13},\quad
\widetilde{\theta}_{23}=\frac{\pi}{2}-\theta_{23},\quad
\widetilde{\delta}=\pi-\delta
\end{equation}
One can also prove that two neutrino mass matrices related by $\mu-\tau$ permutation transformation share the same neutrino mass
eigenvalues.  It is straightforward to verify that such a permutation symmetry exists between
\begin{equation}\begin{split}
&P1\leftrightarrow P2\quad\quad P3\leftrightarrow P5\quad\quad P4\leftrightarrow P6\quad\quad P7\leftrightarrow P7\quad\quad P8\leftrightarrow P9\quad\quad  P10\leftrightarrow P11\\
&P12\leftrightarrow P12\quad\quad P13\leftrightarrow P13 \quad\quad P14\leftrightarrow P15
\end{split}\end{equation}
Notice that the pattern P7, P12 and P13 transform into themselves under the $\mu-\tau$ permutation.
Thus among fifteen texture patterns we studied, only nine of them is independent

We present the allowed range of oscillation parameters for each viable patterns in Table.III and Fig.1-15.  In the figures, the light blue
bands represent the 1$\sigma$ uncertainty in determination of
$\theta_{12}, \theta_{23}$ and $\theta_{13}$ while they plus the
beige bands correspond to the 2$\sigma$
uncertainty. Some interesting observations are summarized as follow
\begin{itemize}
\item Among the fifteen possible patterns, the P7,P10 and P11 patterns are excluded by the current experimental data at 3$\sigma$ confidence level. The P4 pattern,  though allowed at $3\sigma$ level, is excluded at $2\sigma$ level. 
\item The P1, P2 and P12 patterns are phenomenologically allowed for both NO and IO spectrum of neutrino masses.
The P4 and P6 patterns are phenomenologically allowed only for NO spectrum of neutrino masses while P3, P5,P8,P9, P13, P14 and P15 patterns are phenomenologically allowed only for IO spectrum of neutrino masses. 

\item The P4 pattern predicts $\theta_{23}>45^{\circ}$ while for P6 we get $\theta_{23}<45^{\circ}$. In particular, the allowed region of $\theta_{23}$ from P12 pattern is tightly located at around $45^{\circ}$. Taking the NO spectrum as an example, we derive the analytical approximate formula up to the second order of $\sin\theta_{13}$
\begin{equation}
\xi=\frac{m_{2}}{m_{3}} \approx \frac{1}{\cos \theta_{12}}\big(1+\tan \theta_{12} \tan2\theta_{23} \sin\theta_{13}-\frac{1}{4}\tan^2\theta_{12}\tan^{2}\theta_{23}\sin^{2}\theta_{13}\big)
\label{mr1}\end{equation}
where, without loss of generality, we have set $\delta=0$. Clearly, the smallness of mass ratio $\xi\approx\sqrt{\delta m^{2}/\Delta m^{2}}$ implies a large cancellation in Eq.\eqref{mr1} which, as shown in Fig. 5, only happens when $\theta_{23}$ approaches $\pi/4$ enough. Thus the two highly constrained spaces for $\theta_{23}$ besides $45^{\circ}$ can be regarded as the two solutions from Eq.\eqref{mr1}.
One can apply the similar analysis to IO spectrum of neutrino masses, where in leading order of $\sin\theta_{13}$ we have
\begin{equation}
\zeta=\frac{m_{2}}{m_{1}}\approx \tan^{2}\theta_{12}\Big(1+\frac{2\tan2\theta_{23}\cos\delta\sin\theta_{13}}{\sin\theta_{12}\cos\theta_{12}}\Big)
\end{equation}
One can see that the value of $\theta_{23}$ should be located at around $\pi/4$ in order to obtain the mass ratio $\zeta\approx 1$, 

\item Although all the viable patterns admit a large Dirac CP violation, the P8, P9, P10 P11, P14 and P15 patterns require a Dirac CP-violating phase $\delta$ highly located at around $90^{\circ}(270^{\circ})$, leading to a large Dirac CP-violating effect $J_{CP}>0.02$. If the Dirac CP-violation is finally observed to be very small, these patterned can be excluded.  We mention that the latest analysis gives a best-fit value of $1.38\pi$ for NO spectrum and $1.31\pi$ for IO spectrum\cite{Capozzi:2017ipn}, which strengthens the trend in favor of $\delta\sim 3\pi/2$. 

\item The viable textures with NO spectrum of neutrino masses predict tiny $m_{ee}$ (at order of $0.001$eV), rendering it very challenging to be detected in future $0\nu\nu\beta$ decay experiments. This is due to the fact that, in the scenario of $m_{1}=0, m_{2}=\sqrt{\delta m^{2}}\ll m_{3}$, one easily obtains $m_{ee}\approx \sqrt{\Delta m^2}s_{13}^{2}$  from Eq.\eqref{mee}, with the value of $m_{ee}$ largely suppressed by small $s_{13}^{2}$ .  For IO spectrum of neutrino masses, on the other hand, each viable texture can predict $m_{ee}$ in the order of $0.01$eV, which is promising to be detected in the forthcoming experiments.
\end{itemize}

\begin{center}
\begin{figure}
\includegraphics[scale=0.49]{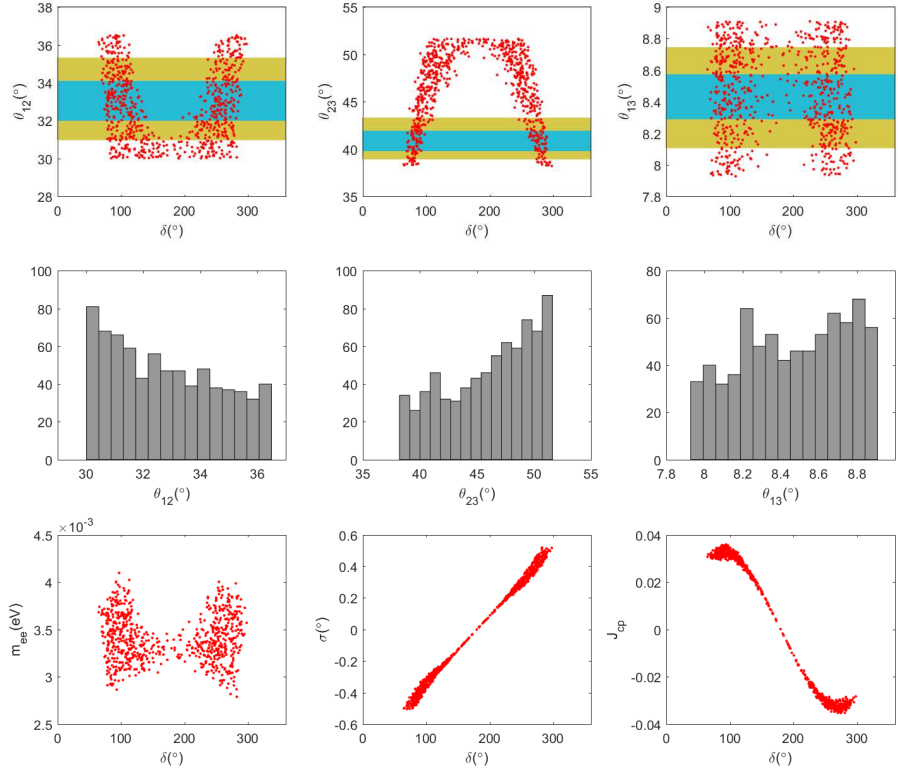}
\setlength{\abovecaptionskip}{-0.8cm}
\caption {The correlation plots for NO P1 pattern. }\label{p1no}
\end{figure}
\end{center}

\begin{center}
\begin{figure}
\includegraphics[scale=0.49]{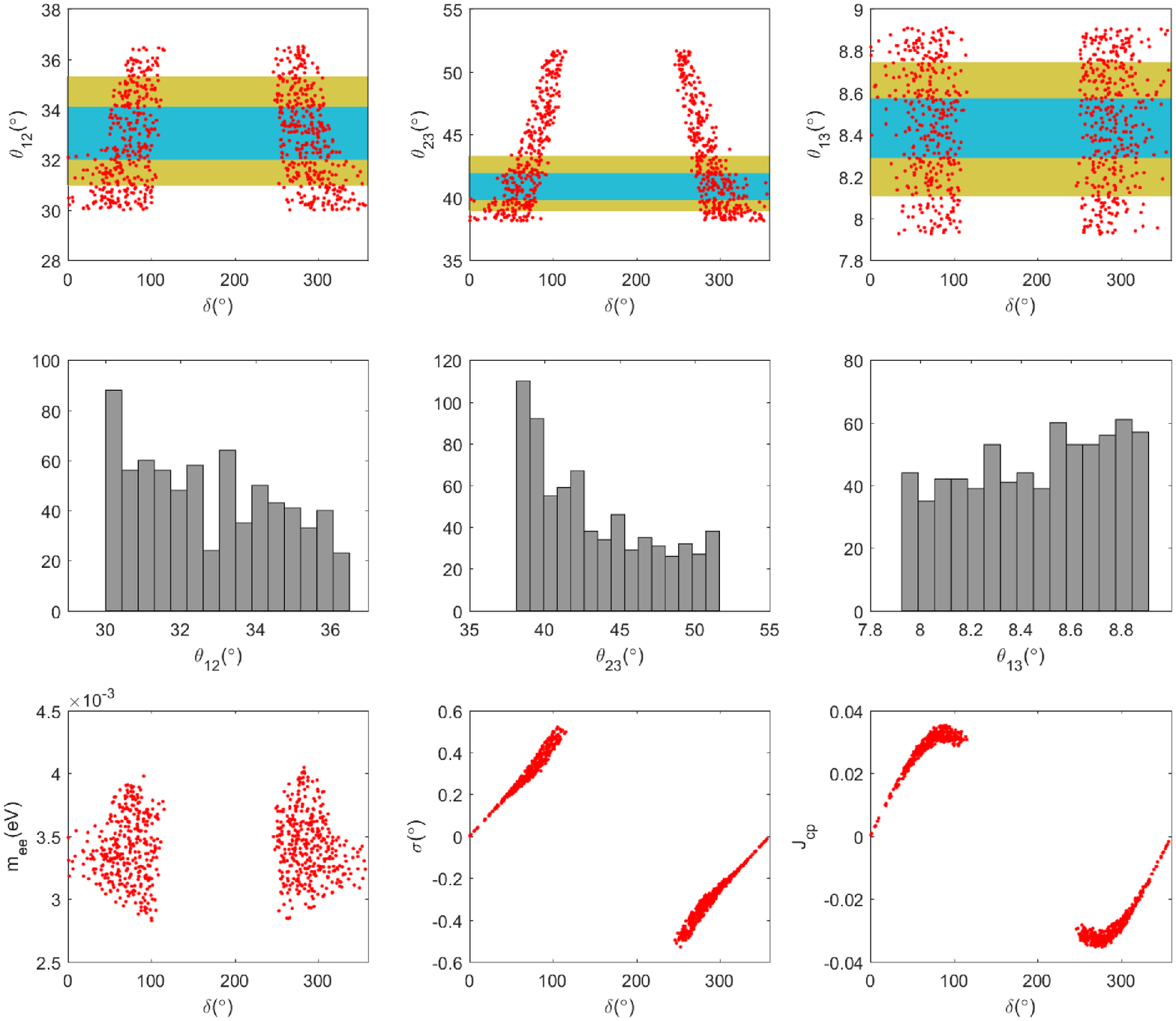}
\setlength{\abovecaptionskip}{-0.8cm}
\caption {The correlation plots for NO P2 pattern. }\label{p2no}
\end{figure}
\end{center}

\begin{figure}
\includegraphics[scale=0.49]{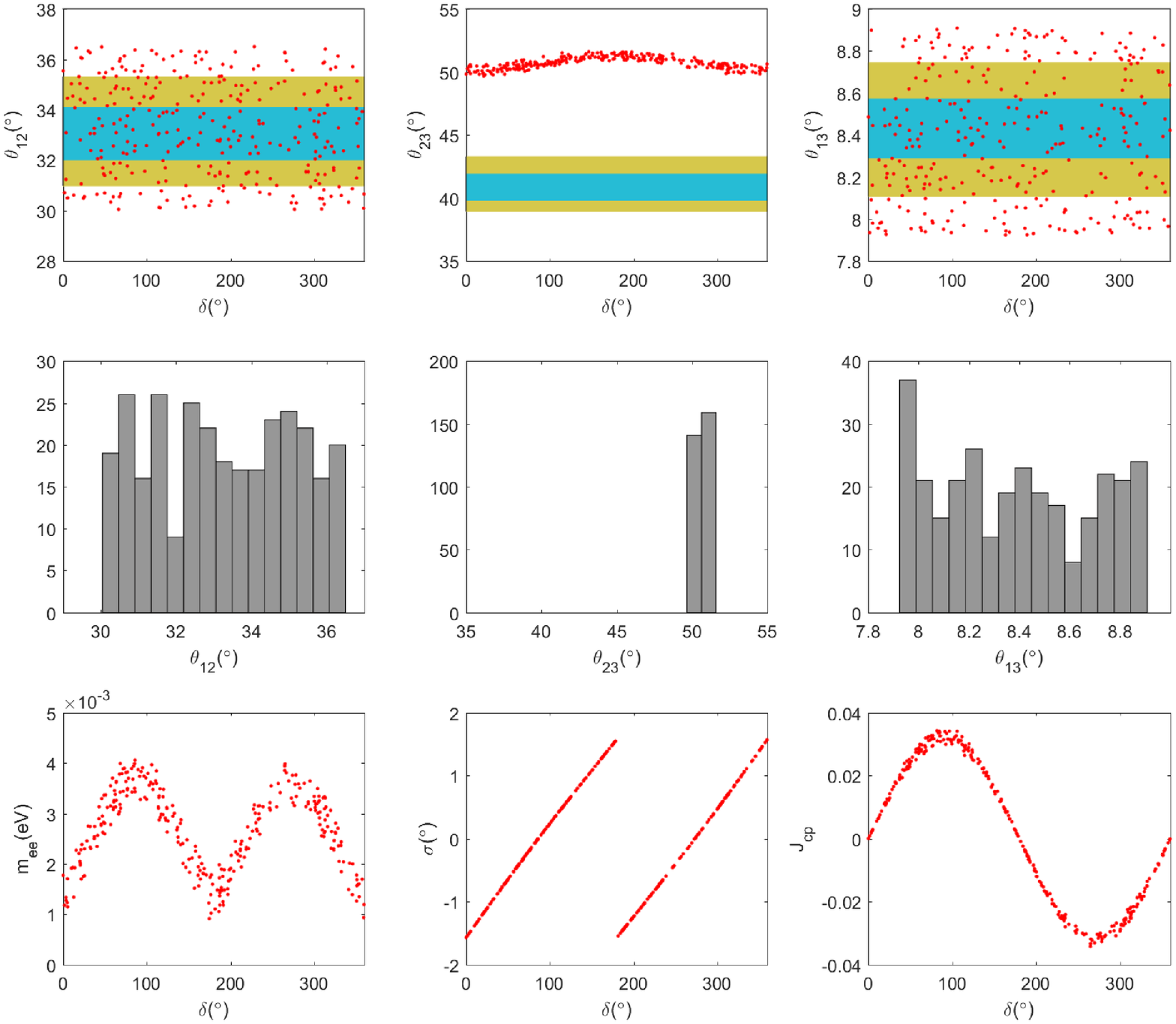}
\setlength{\abovecaptionskip}{-0.8cm}
\caption {The correlation plots for NO P4 pattern. }\label{p4no}
\end{figure}

\begin{figure}
\includegraphics[scale=0.49]{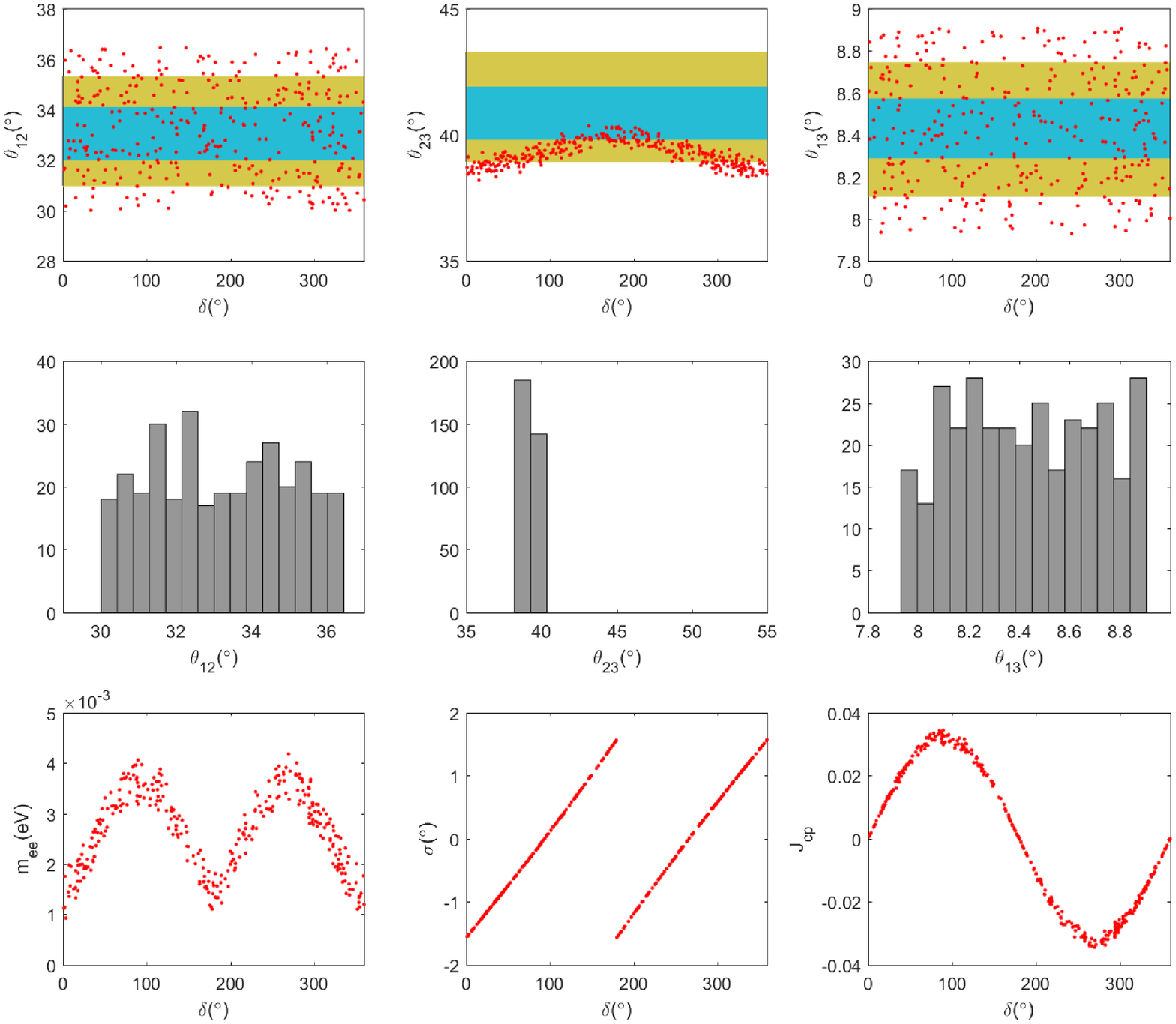}
\setlength{\abovecaptionskip}{-0.8cm}
\caption {The correlation plots for NO P6 pattern.}\label{p6no}
\end{figure}

\begin{figure}
\includegraphics[scale=0.49]{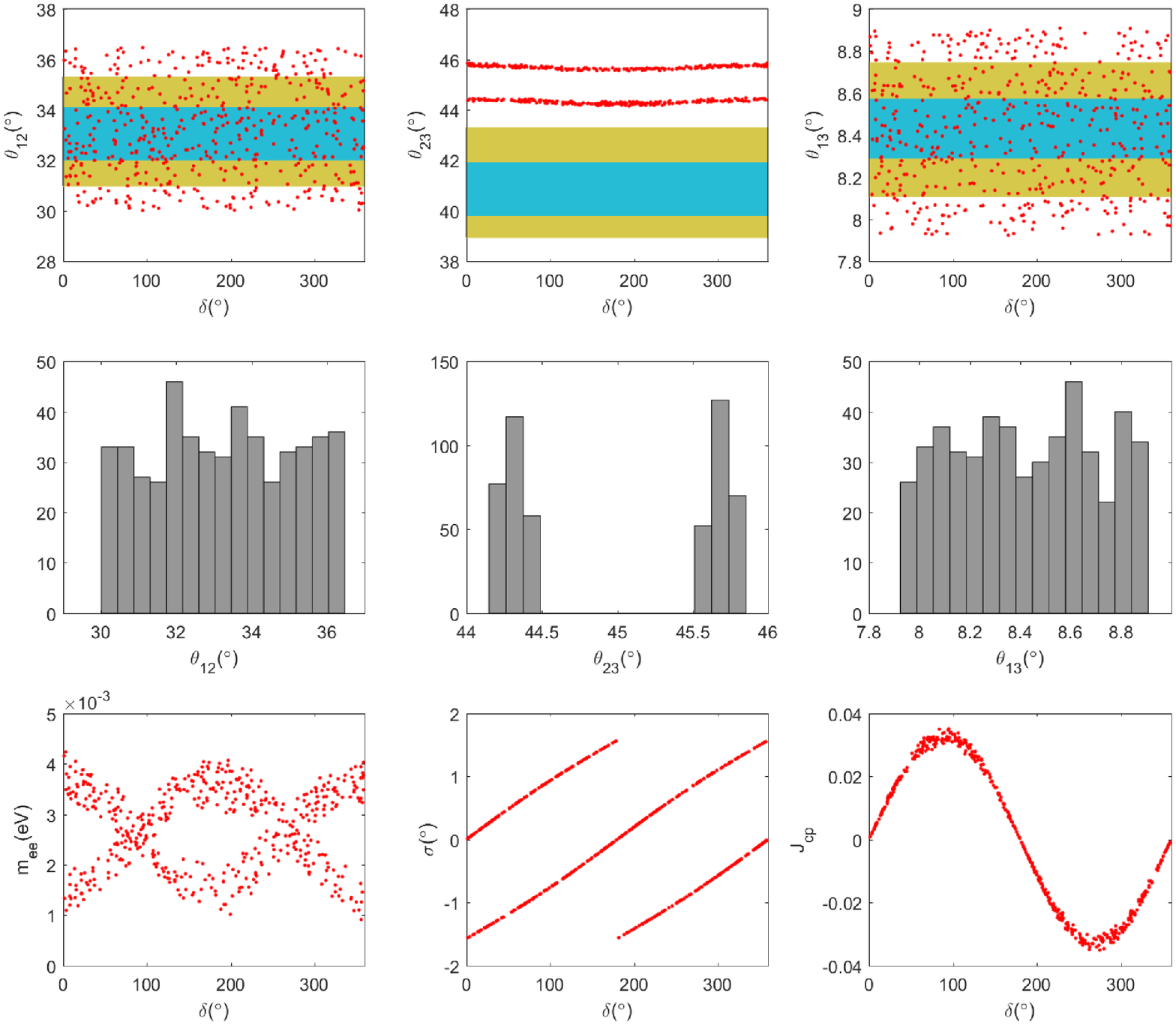}
\setlength{\abovecaptionskip}{-0.8cm}
\caption {The correlation plots for NO P12 pattern.}\label{p12no}
\end{figure}

\begin{figure}
\includegraphics[scale=0.49]{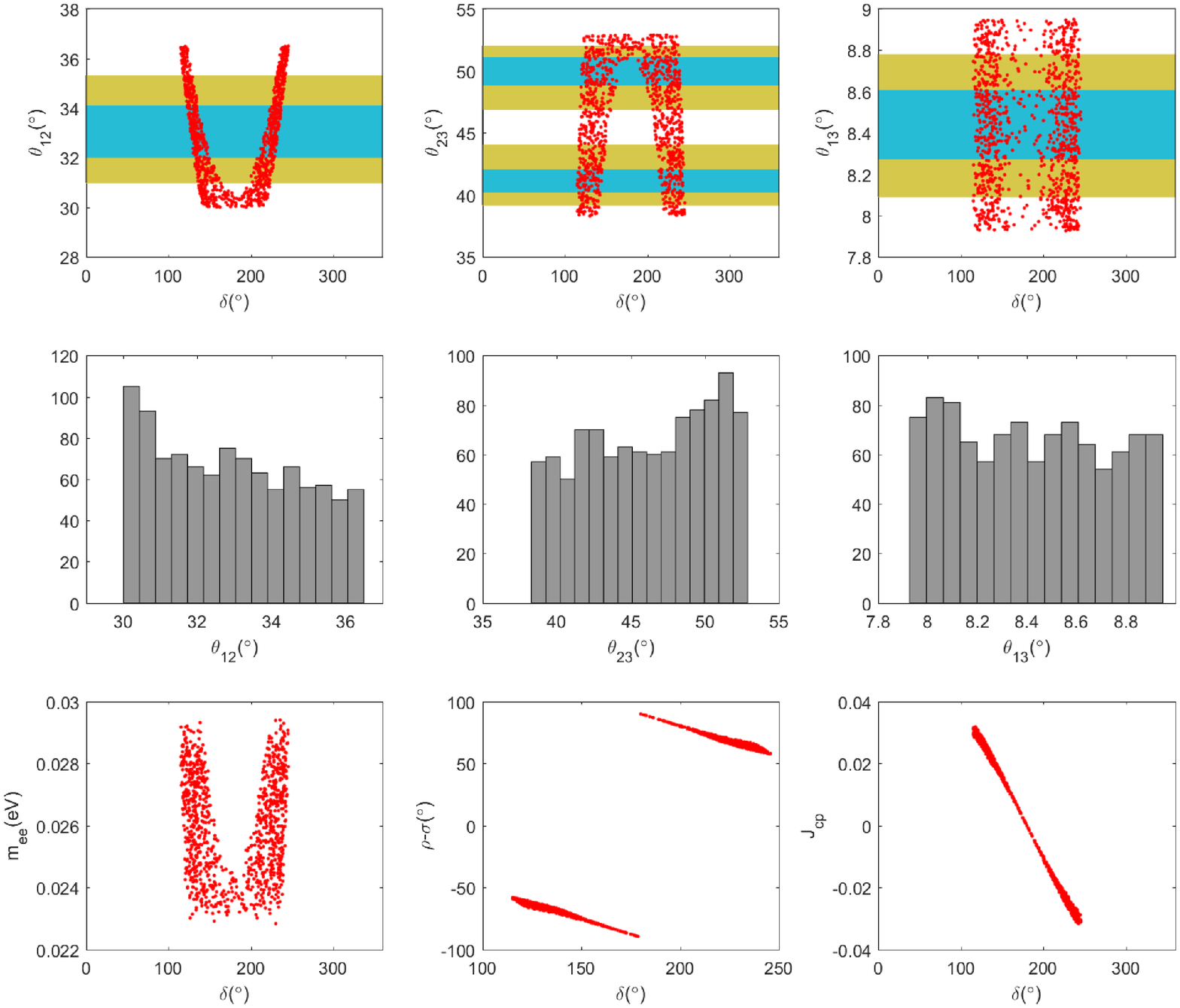}
\setlength{\abovecaptionskip}{-0.8cm}
\caption {The correlation plots for IO P1 pattern.}\label{p1io}
\end{figure}

\begin{figure}
\includegraphics[scale=0.49]{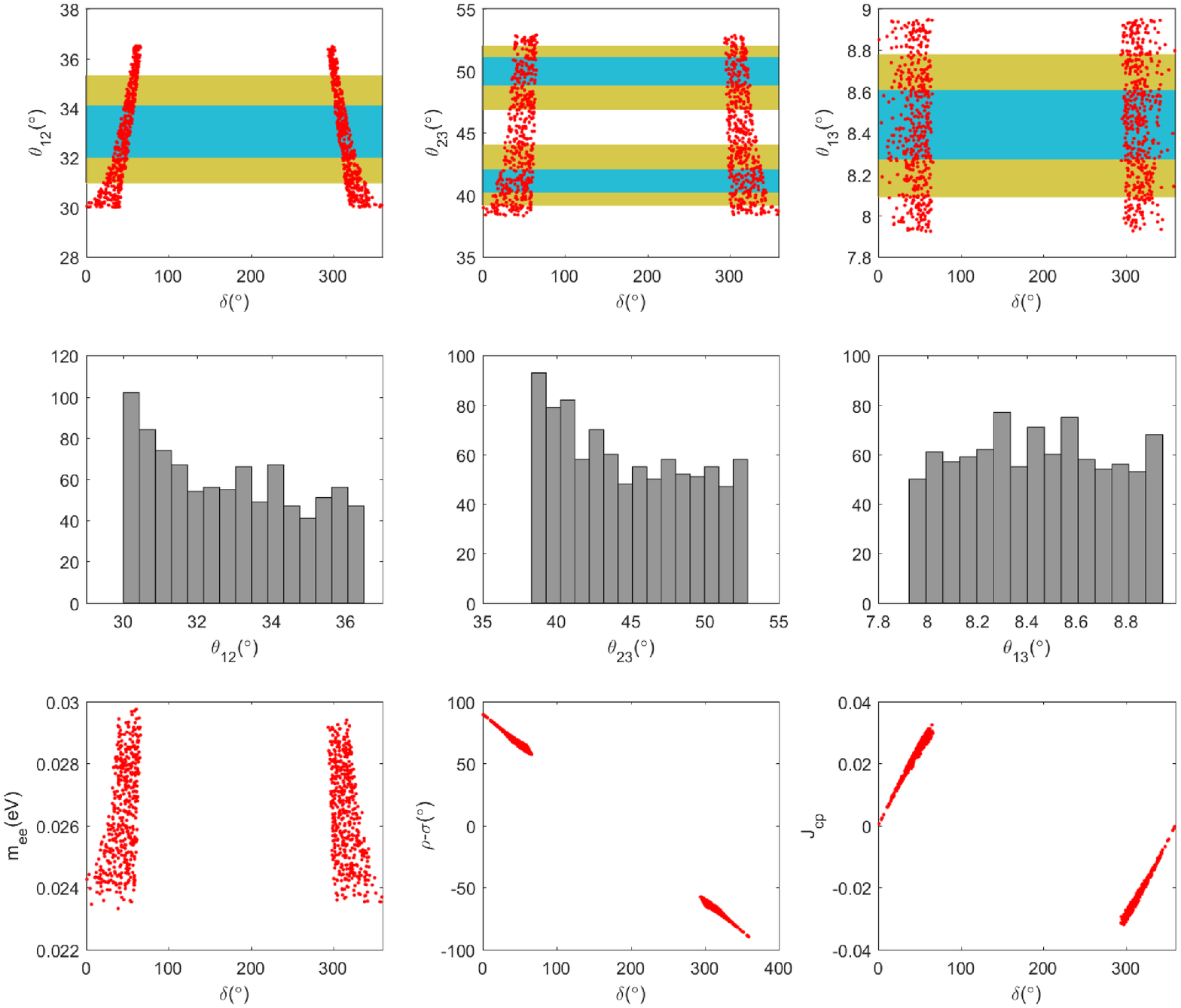}
\setlength{\abovecaptionskip}{-0.8cm}
\caption {The correlation plots for IO P2 pattern. }\label{p2io}
\end{figure}

\begin{figure}
\includegraphics[scale=0.49]{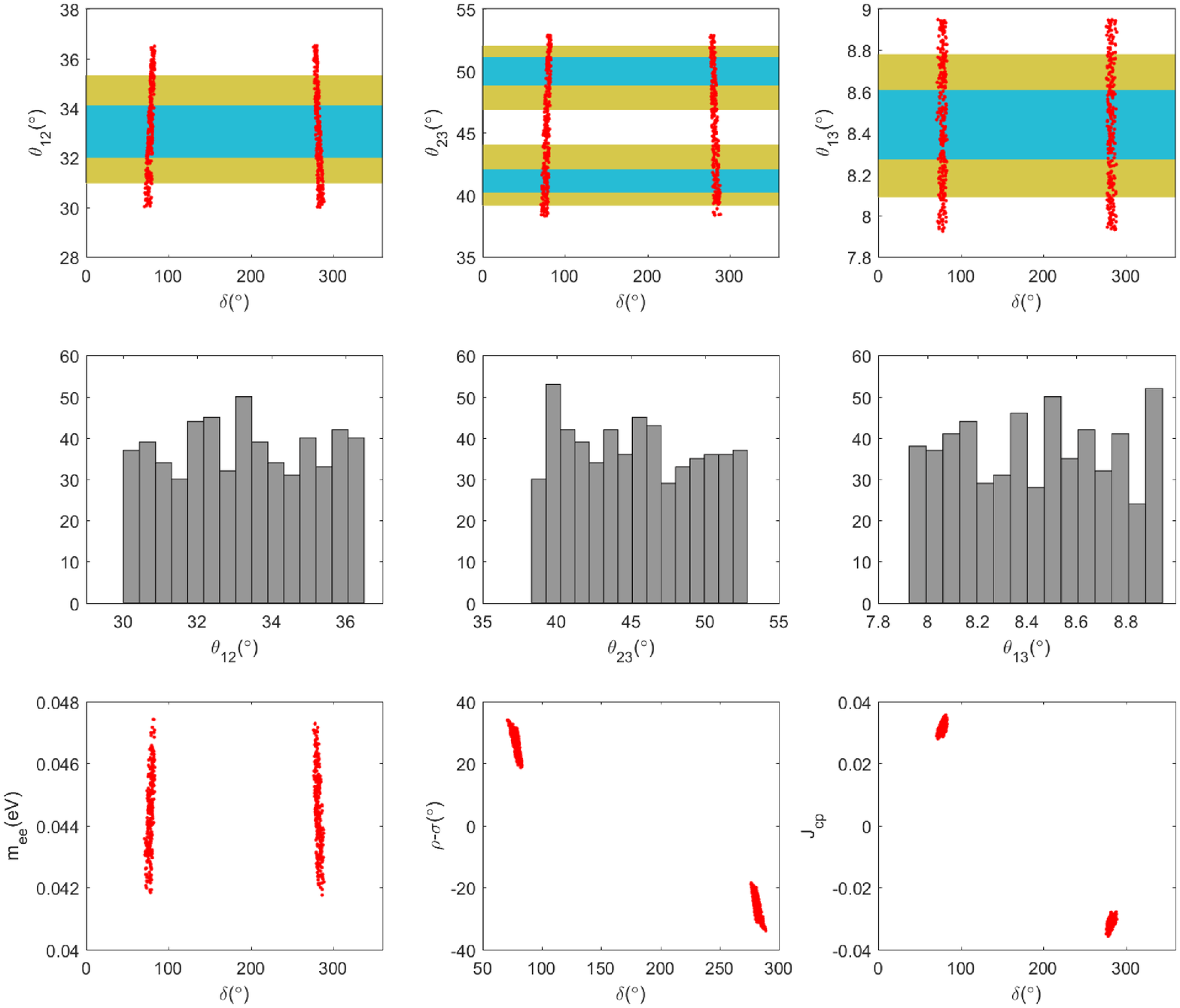}
\setlength{\abovecaptionskip}{-0.8cm}
\caption {The correlation plots for IO P3 pattern.}\label{p3io}
\end{figure}

\begin{figure}
\includegraphics[scale=0.49]{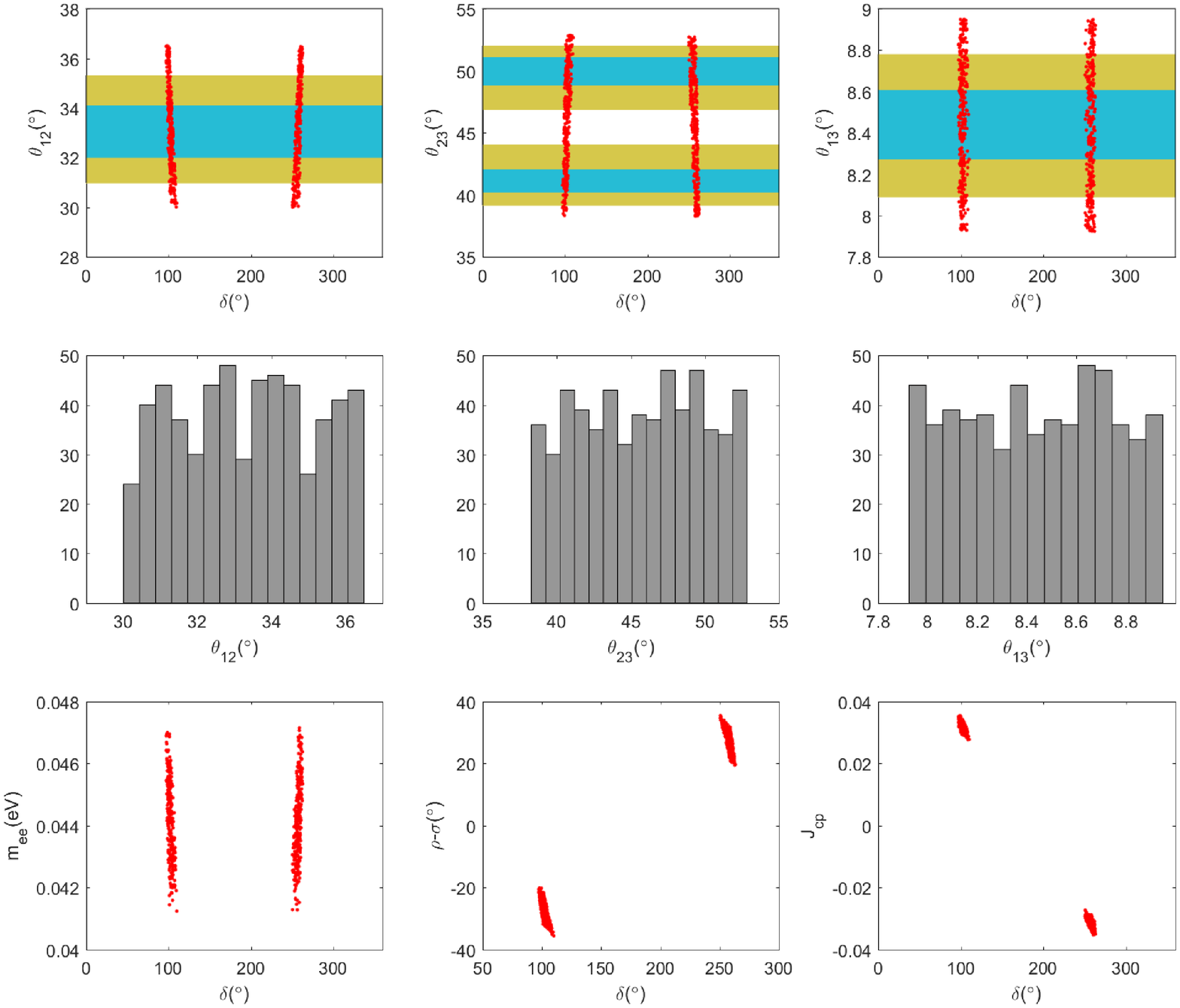}
\setlength{\abovecaptionskip}{-0.8cm}
\caption {The correlation plots for IO P5 pattern. }\label{p4io}
\end{figure}

\begin{figure}
\includegraphics[scale=0.49]{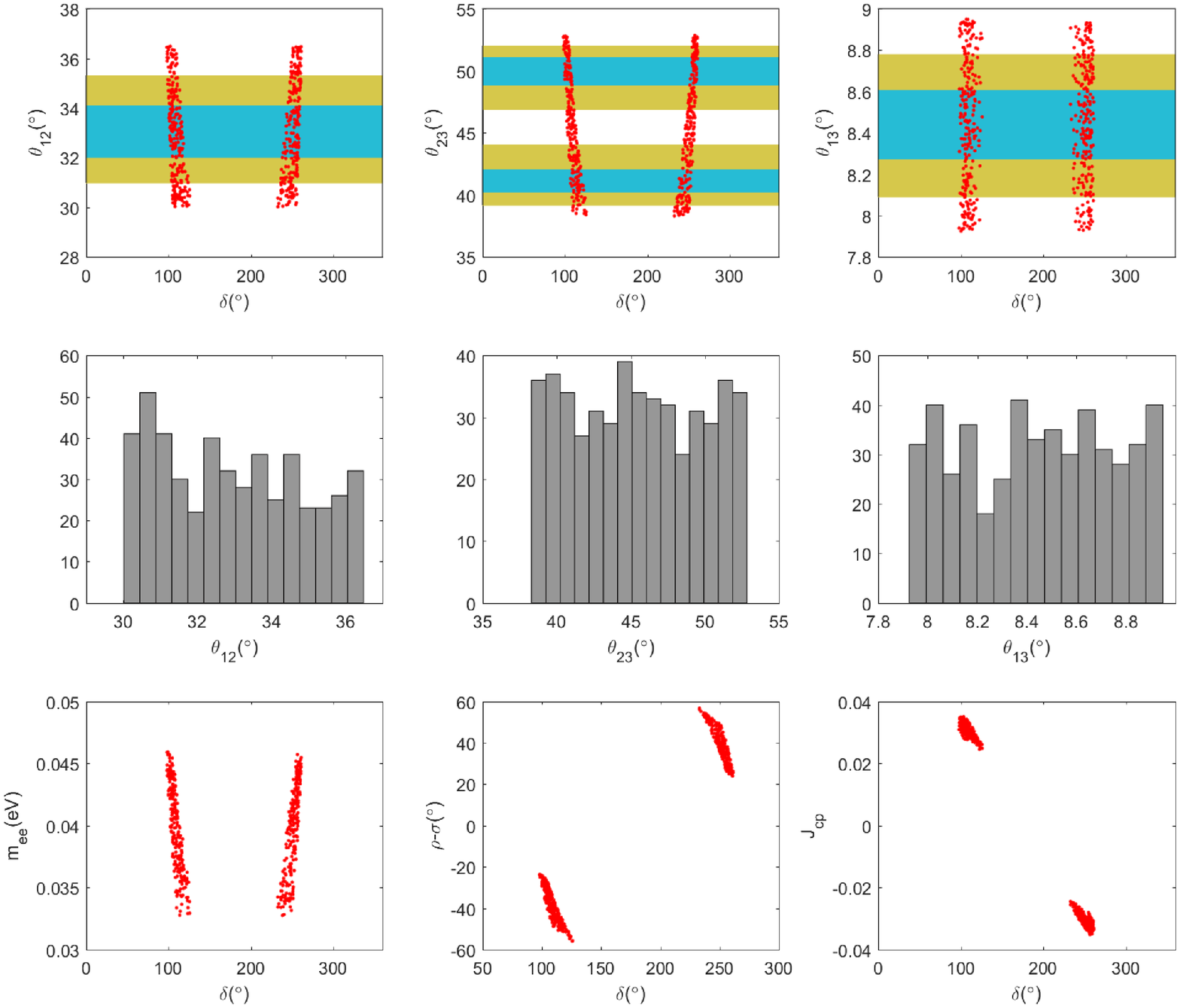}
\setlength{\abovecaptionskip}{-0.8cm}
\caption {The correlation plots for IO P8 pattern. }\label{p8io}
\end{figure}
\begin{figure}
\includegraphics[scale=0.49]{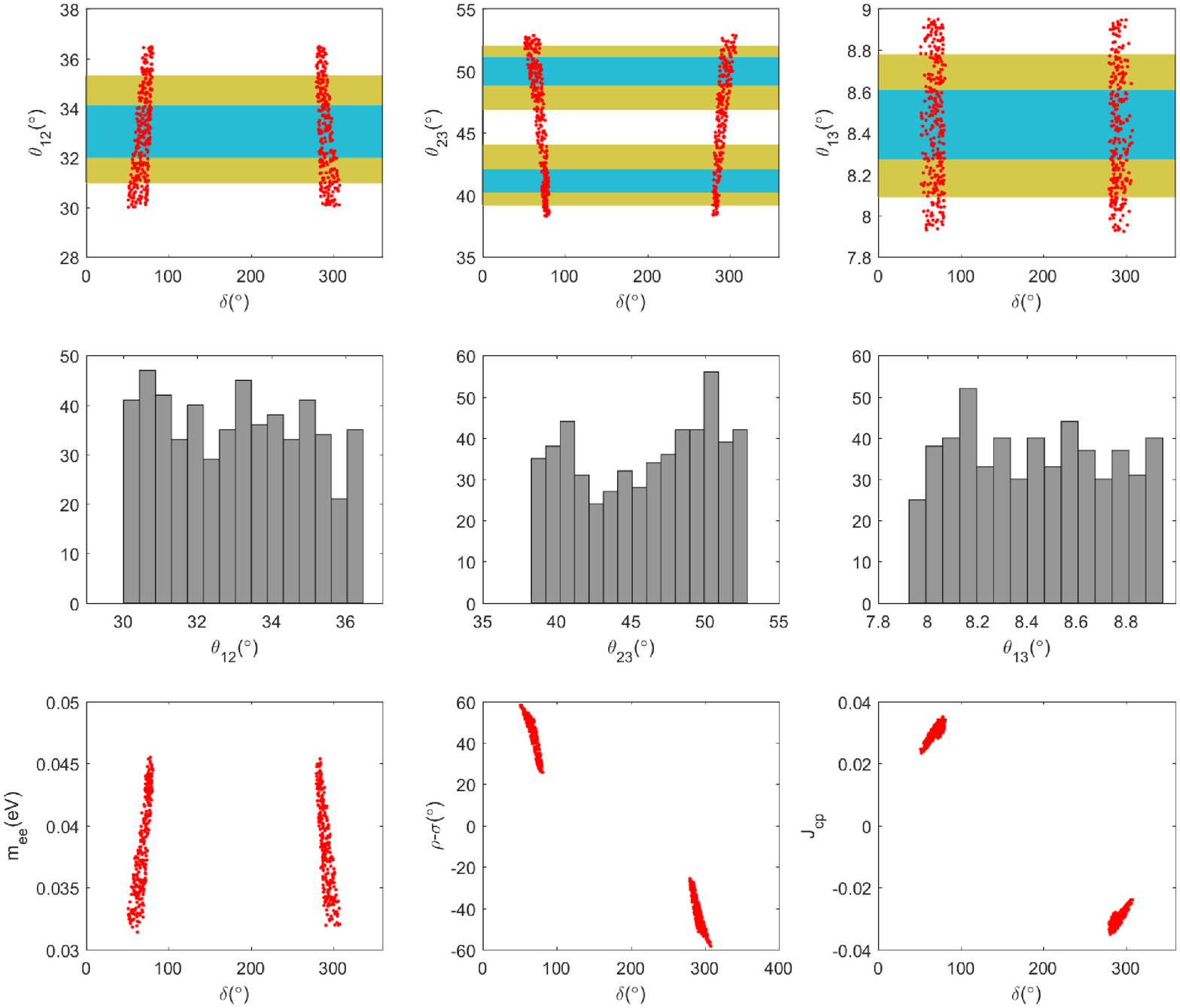}
\setlength{\abovecaptionskip}{-0.8cm}
\caption {The correlation plots for IO P9 pattern.}\label{p9io}
\end{figure}

\begin{figure}
\includegraphics[scale=0.49]{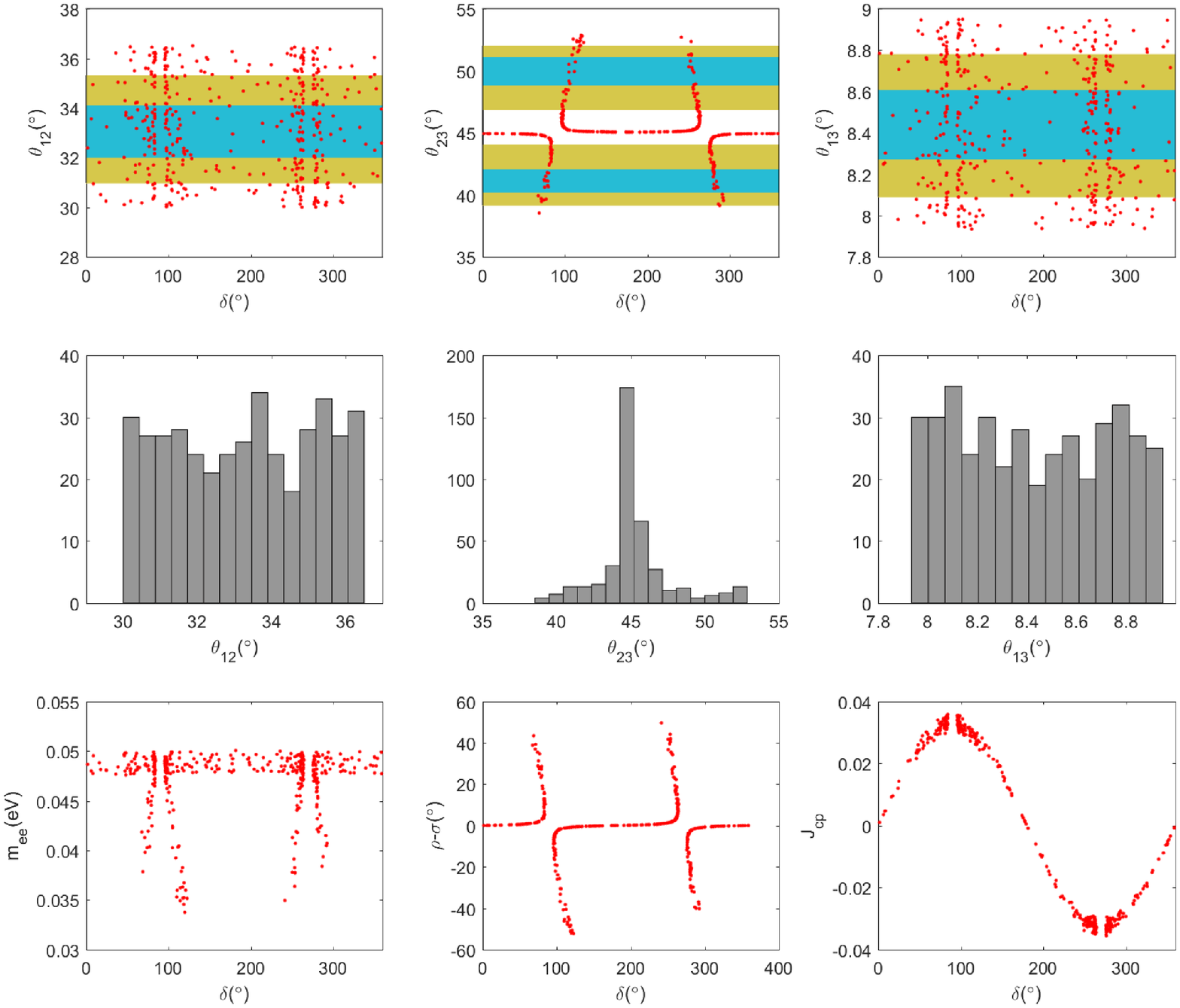}
 \setlength{\abovecaptionskip}{-0.8cm}
\caption {The correlation plots for IO P12 pattern. }\label{p12io}
\end{figure}

\begin{figure}
\includegraphics[scale=0.49]{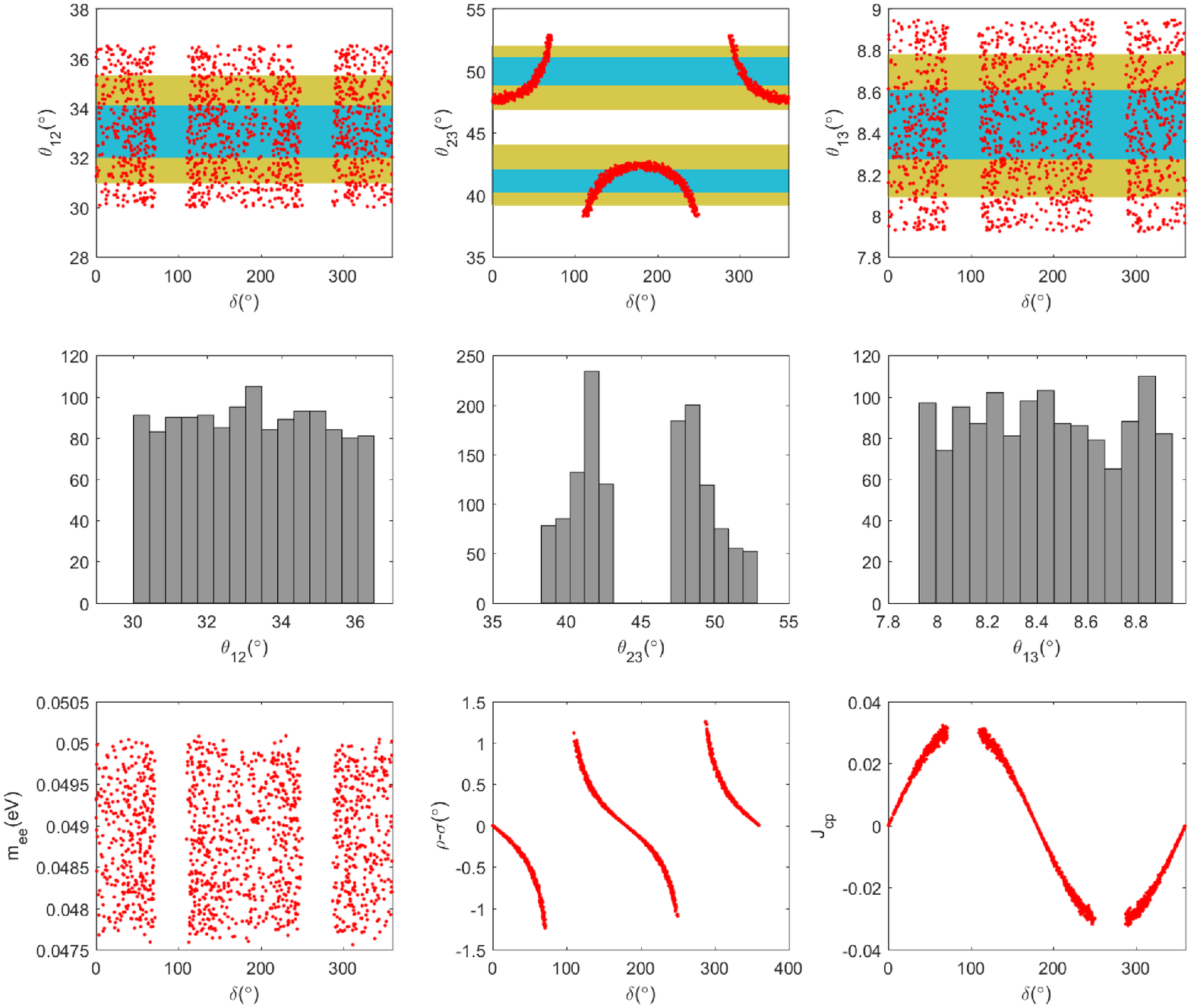}
\setlength{\abovecaptionskip}{-0.8cm}
\caption {The correlation plots for IO P13 pattern }\label{p13io}
\end{figure}
\begin{figure}
\includegraphics[scale=0.49]{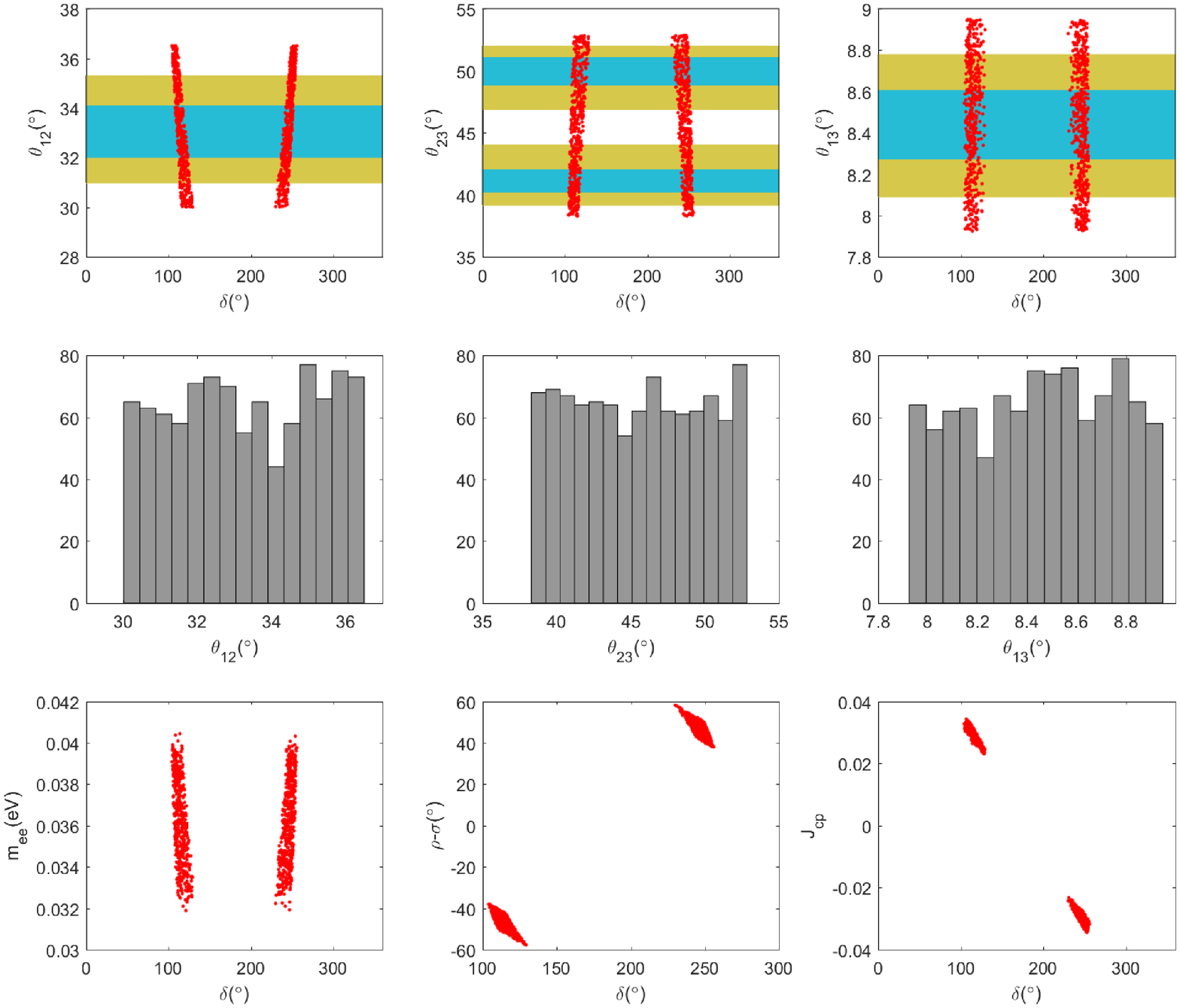}
\setlength{\abovecaptionskip}{-0.8cm}
\caption {The correlation plots for IO P14 pattern. }\label{p14io}
\end{figure}

\begin{figure}
\includegraphics[scale=0.49]{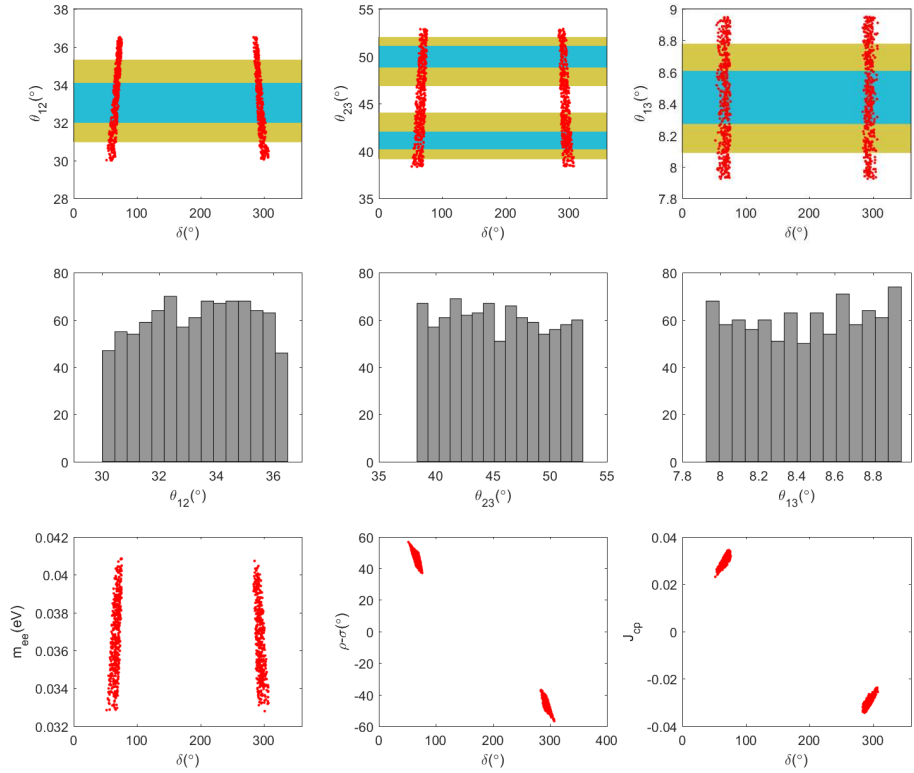}
\setlength{\abovecaptionskip}{-0.8cm}
\caption {The correlation plots for IO P15 pattern. }\label{p15io}
\end{figure}

\section{Theoretical realization for a concrete model}
In this section, we present a detailed illustration on how flavor symmetry gives rise to the desired texture structure. The neutrino mass matrices with texture equalities have been realized by using a non-Abelian flavor symmetry e.g $S_{3}$\cite{hybrid2} or $A_{4}$\cite{co}. On the other hand, one notices that as good approximation of vanishing neutrino mass, a ultralight neutrino masse implies a large mass hierarchy between three generations of neutrinos. To realize  hierarchies between neutrino masses, a popular approach is the Froggatt-Nielsen mechanism\cite{FN} that was originally developed for quark sector while is also suitable to understand the lepton mass and mixing pattern. In the following, we just take $P12$ pattern as a concrete example, although a complete realization for other patterns is also interesting and necessary. 

We extend the particle content of Standard Model(SM) with three right-handed neutrinos $\nu_{Ri}(i=1,2,3)$, three SM-like  doublet scalars $H_{i}(i=1,2,3)$ responsible for charged lepton masses generation, four additional doublet scalars $\phi_{i}(i=1,2,3,4)$ responsible for Dirac neutrino mass matrix $M_{D}$, two singlet scalars $\sigma_{i}(i=1,2)$ giving rise to the right-handed neutrino mass matrix $M_{R}$. Then the neutrino mass matrix is obtained by the canonical type-I seesaw formula $M_{\nu}=M_{D}M_{R}^{-1}M_{D}$. The general Lagrangian that producing to lepton masses is then given by
\begin{equation}
 \mathcal{L}=\Big(\frac{ \left \langle\Phi\right\rangle}{\Lambda}\Big)^{Q_{L_{i}}+Q_{l_{Rj}}}y_{ij}^{(k)}\bar{L}_{i}H_{k}l_{Rj}+
 \Big(\frac{ \left \langle\Phi\right\rangle}{\Lambda}\Big)^{Q_{L_{i}}+Q_{\nu_{Rj}}}h_{ij}^{(k)}\bar{L}_{i}\widetilde{\phi}_{k}^{\prime}\nu_{Rj} +\Big(\frac{ \left \langle\Phi\right\rangle}{\Lambda}\Big)^{Q_{\nu_{Ri}}+Q_{\nu_{Rj}}}g_{ij}^{(k)}\sigma_{k}\nu_{Ri}\nu_{Rj}+h.c
\end{equation}
where $\widetilde{\phi}_{k}^{\prime}\equiv i\sigma_{2}\phi_{k}^{\prime}$ and we denote $ \epsilon\equiv\left \langle\Phi\right\rangle/\Lambda$.
The $Q_{\alpha} $($\alpha = L,l_{R},\nu_{R}$) are interpreted as the FN charges for SM fermion ingredients under which different generations may be charged differently. The flavon $\Phi$ obtains the vaccum (VEV) 
$\left\langle\Phi\right\rangle$ that  breaks the FN symmetry. We assign the FN charges for lepton sector as 
\begin{equation}\begin{split}
\bar L_{1,2,3}: (a+1,a,a) \\
l_{R1,2,3}: (0,1,2)\\
\nu_{R1,2,3}: (d,c,b)
\end{split}\end{equation}
The realistic mass hierarchy between three charged leptons are successfully reproduced by adopting the assignment of FN charges for $L_{i}$ and $l_{RI}$ with $\epsilon\simeq 0.05-0.1$\cite{Sato:2000kj}. The explicit forms of Dirac neutrino mass matrix $M_{D}$ and right-handed neutrino mass matrix $M_{R}$ are
\begin{equation}\begin{split}
M_{D}=\sum_{k=1}^3 M_{D}^{(k)}&=\sum_{k=1}^3 v_{k}\left(\begin{array}{ccc}
\epsilon^{a+1}&0&0\\
0&\epsilon^{a}&0\\
0&0&\epsilon^{a}
\end{array}\right)
\left(\begin{array}{ccc}
h_{11}^{(k)}&h_{12}^{(k)}&h_{13}^{(k)}\\
h_{21}^{(k)}&h_{22}^{(k)}&h_{23}^{(k)}\\
h_{31}^{(k)}&h_{32}^{(k)}&h_{33}^{(k)}
\end{array}\right)
\left(\begin{array}{ccc}
\epsilon^{d}&0&0\\
0&\epsilon^{c}&0\\
0&0&\epsilon^{b}
\end{array}\right)\\
&=\sum_{k=1}^3v_{k}\left(\begin{array}{ccc}
h_{11}^{(k)}\epsilon^{d+a+1}&h_{12}^{(k)}\epsilon^{c+a+1}&h_{13}^{(k)}\epsilon^{b+a+1}\\
h_{21}^{(k)}\epsilon^{d+a}&h_{22}^{(k)}\epsilon^{c+a}&h_{23}^{(k)}\epsilon^{b+a}\\
h_{31}^{(k)}\epsilon^{d+a}&h_{32}^{(k)}\epsilon^{c+a}&h_{33}^{(k)}\epsilon^{b+a}
\end{array}\right)
\end{split}\end{equation}
and 
\begin{equation}\begin{split}
M_{R}=\sum_{k=1}^2 M_{R}^{(k)}=&\sum_{k=1}^2 v_{k}^{\prime}\left(\begin{array}{ccc}
\epsilon^{d}&0&0\\
0&\epsilon^{c}&0\\
0&0&\epsilon^{b}
\end{array}\right)
\left(\begin{array}{ccc}
g_{11}^{(k)}&g_{12}^{(k)}&g_{13}^{(k)}\\
g_{21}^{(k)}&g_{22}^{(k)}&g_{23}^{(k)}\\
g_{31}^{(k)}&g_{32}^{(k)}&g_{33}^{(k)}
\end{array}\right)
\left(\begin{array}{ccc}
\epsilon^{d}&0&0\\
0&\epsilon^{c}&0\\
0&0&\epsilon^{b}
\end{array}\right)\\
&=\sum_{k=1}^2 v_{k}^{\prime}
\left(\begin{array}{ccc}
g_{11}^{(k)}\epsilon^{2d}&g_{12}^{(k)}\epsilon^{d+c}&g_{13}^{(k)}\epsilon^{d+b}\\
g_{21}^{(k)}\epsilon^{d+c}&g_{22}^{(k)}\epsilon^{2c}&g_{23}^{(k)}\epsilon^{c+b}\\
g_{31}^{(k)}\epsilon^{d+b}&g_{32}^{(k)}\epsilon^{c+b}&g_{33}^{(k)}\epsilon^{2b}
\end{array}\right)
\end{split}\end{equation}
Following the spirit of Ref.\cite{Lashin:2013xha}, we consider a cooperation of $\mu-\tau$ permutation symmetry and $Z_{8}$ horizontal symmetry i.e $S_{\mu-\tau} \times Z_{8}$ under which the relevant particle fields transform as
\begin{equation}\begin{split}
&(L_{e},L_\mu,L_\tau)\xrightarrow{S}(L_{e},L_\mu,L_\tau),\quad (e_{R},\mu_R,\tau_R)\xrightarrow{S}(e_{R},\tau_R,\mu_R),\quad (H_{1},H_{2},H_{3}) \xrightarrow{S}  (H_{1},H_{3},H_{2})\\
&(\nu_{R1},\nu_{R2},\nu_{R3})\xrightarrow{S} (\nu_{R1},\nu_{R3},\nu_{R2}),\quad (\phi_{1},\phi_{2},\phi_{3},\phi_{4}) \xrightarrow{S}(\phi_{1},\phi_{2},\phi_{3},-\phi_{4}),\quad (\sigma_{1},\sigma_{2})\xrightarrow{S}(\sigma_{1},\sigma_{2})\\
&(L_{e},L_\mu,L_\tau)\xrightarrow{Z_{8}}(L_{e},-L_\mu,-L_\tau),\quad  (e_{R},\mu_R,\tau_R)\xrightarrow{Z_{8}}(e_{R},\tau_R,-\mu_R),\quad  (H_{1},H_{2},H_{3}) \xrightarrow{Z_{8}}  (H_{1},-H_{2},-H_{3})\\
&(\nu_{R1},\nu_{R2},\nu_{R3})\xrightarrow{Z_{8}} (\omega\nu_{R1},\omega^{3}\nu_{R3},\omega^{3}\nu_{R2}),\quad
(\phi_{1},\phi_{2},\phi_{3},\phi_{4}) \xrightarrow{Z_{8}}(\omega\phi_{1},\omega^{3}\phi_{2},\omega^{7}\phi_{3},\omega^{3}\phi_{4})\\
&(\sigma_{1},\sigma_{2})\xrightarrow{Z_{8}}(\omega^{6}\sigma_{1},\omega^{2}\sigma_{2})
\end{split}\end{equation}
As proposed in Ref.\cite{Lashin:2013xha}, one can further assume a hierarchy in $H_{i}$'s vacuums ($ \left \langle H_{3}\right\rangle\gg  \left \langle H_{1}\right\rangle, \left \langle H_{2}\right\rangle$). Then the $S_{\mu-\tau}\times Z_{8}$-invariant Lagrangian relevant to the charged lepton sector leads to the texture structure of $M_{l}M^{\dagger}_{l}$ given by
\begin{equation}
M_{l}M^{\dagger}_{l}=\left(\begin{array}{ccc}
A^2&0&0\\
0&D^2&DC\\
0&DC&C^2
\end{array}\right)
\end{equation}
with 
\begin{equation}
\frac{B}{C}\sim \frac{m_{e}}{m_{\mu}}=2.8\times 10^{-4}\quad \frac{D}{C}\sim \frac{m_{\mu}}{m_{\tau}}=5.9\times 10^{-2}
\end{equation}
As required, we construct the model in the flavor basis corrected by an extremely small rotation angle less than $10^{-2}$ 

Now move to the neutrino sector. The $S_{\mu-\tau}\times Z_{8}$ symmetry imposes $b=c$ and the
form of the neutrino mass matrices as follows:
\begin{equation}\begin{split}
&M_{D}^{(1)}=  \left(\begin{array}{ccc}
A_{1}\epsilon^{d+a+1}&0&0\\
0&0&0\\
0&0&0
\end{array}\right)\quad
M_{D}^{(2)}= \left(\begin{array}{ccc}
0&B_{2}\epsilon^{c+a+1}&-B_{2}\epsilon^{c+a+1}\\
0&0&0\\
0&0&0
\end{array}\right)\\
&M_{D}^{(3)}= \left(\begin{array}{ccc}
0&0&0\\
0&C_{3}\epsilon^{c+a}&D_{3}\epsilon^{c+a}\\
0&D_{3}\epsilon^{c+a}&C_{3}\epsilon^{c+a}
\end{array}\right)\quad
M_{D}^{(4)}= \left(\begin{array}{ccc}
0&B_{4}\epsilon^{c+a+1}&B_{4}\epsilon^{c+a+1}\\
0&0&0\\
0&0&0
\end{array}\right)
\end{split}\end{equation}
where
\begin{equation}
A_{1}\equiv v_{1}h_{11}^{(1)},\quad B_{2}\equiv v_{2}h_{12}^{(2)}=-v_{2}h_{13}^{(2)}\quad C_{3}\equiv v_{3} h_{22}^{(3)}=v_{3}h_{33}^{(3)}\quad  D_{4}\equiv v_{4}h_{23}^{(3)}=v_{4}h_{32}^{(3)}
\end{equation}
and 
\begin{equation}
M_{R}^{(1)}=\left(\begin{array}{ccc}
A_{R}\epsilon^{2d}&0&0\\
0&0&0\\
0&0&0
\end{array}\right)\quad
M_{R}^{(2)}= \left(\begin{array}{ccc}
0&0&0\\
0&C_{R}\epsilon^{2c}&D_{R}\epsilon^{2c}\\
0&D_{R}\epsilon^{2c}&C_{R}\epsilon^{2c}
\end{array}\right)
\end{equation}
where 
\begin{equation}
A_{R}=v_{1}^{\prime}g_{11}^{(1)},\quad C_{R}=v_{2}^{\prime}g_{22}^{(2)}=v_{2}^{\prime}g_{33}^{(2)}, \quad D_{R}=v_{2}^{\prime}g_{23}^{(2)}=v_{2}^{\prime}g_{32}^{(2)}
\end{equation}
Hence the Dirac neutrino mass matrix $M_{D}$ and the right-handed neutrino mass matrix $M_{R}$ are given by
\begin{equation}
M_{D}=\left(\begin{array}{ccc}
A_{D}\epsilon^{d+a+1}&B_{D}^{\prime}\epsilon^{c+a+1}&-B_{D}\epsilon^{c+a+1}\\
0&C_{D}\epsilon^{c+a}&D_{D}\epsilon^{c+a}\\
0&D_{D}\epsilon^{c+a}&C_{D}\epsilon^{c+a}
\end{array}\right)\quad
M_{R}=\left(\begin{array}{ccc}
A_{R}\epsilon^{2d}&0&0\\
0&C_{R}\epsilon^{2c}&D_{R}\epsilon^{2c}\\
0&D_{R}\epsilon^{2c}&C_{R}\epsilon^{2c}
\end{array}\right)
\label{mm}\end{equation}
where we have defined
\begin{equation}\begin{split}
A_{D}\equiv A_{1},\quad & B_{D}^{\prime}\equiv B_{2}+B_{4},\quad  B_{D}\equiv - B_{2}+ B_{4},\quad C_{D}\equiv C_{3},\quad D_{D}\equiv D_{3}
\end{split}\end{equation}

With the help of Eq.\eqref{mm} and seesaw formula $M_{\nu}=M_{D}M_{R}^{-1}M_{D}^{T}$,  the effective neutrino mass matrix $M_{\nu}$ is obtained by a direct calculation, leading to the exact form of  $P_{12}$ pattern i.e
\begin{equation}
M_{\nu}=\left(\begin{array}{ccc}
M_{\nu11}&M_{\nu12}&M_{\nu13}\\
M_{\nu12}&M_{\nu22}&M_{\nu23}\\
M_{\nu13}&M_{\nu23}&M_{\nu22}
\end{array}\right)
\end{equation}

In order to get a ultralight neutrino mass, we adopt the methodology given in Ref.\cite{Fujii:2001zr} . The basic point is to suppose the broken FN symmetry is a discrete $Z_{n}$ symmetry instead of continuous $U(1)_{\text{FN}}$. One further assumes $n=2d$ and $c<d$. Then the  right-handed neutrino mass matrix $M_{R}$ in Eq.\eqref{mm} is 
\begin{equation}
M_{R}=\left(\begin{array}{ccc}
A_{R}&0&0\\
0&C_{R}\epsilon^{2c}&D_{R}\epsilon^{2c}\\
0&D_{R}\epsilon^{2c}&C_{R}\epsilon^{2c}
\end{array}\right)
\end{equation}
We see that In $M_{R}$ the element $A_{R}$ is not suppressed by the power of $\epsilon$ compared with $C_{R}$ and $D_{R}$, which indicates an extremely large right-handed neutrino masses for $N_{1}$ and hence yields an ultralight neutrino mass via type-I seesaw mechanism. The texture hierarchy in the power of $\epsilon$ is derived as
\begin{equation}
M_{\nu}\sim \epsilon^{2a}\left(\begin{array}{ccc}
\epsilon^{2}&\epsilon&\epsilon\\
\epsilon&1&1\\
\epsilon&1&1
\end{array}\right)
\label{ts}\end{equation}
Note that the the FN charges $d$ and $c$ are completely canceled out when applying seesaw formula and $\epsilon^{2a}$ appears as a total factor in the form of $M_{\nu}$. As firstly indicated in Ref.\cite{Sato:2000kj}, the structure of neutrino mass matrices give in Eq.\eqref{ts} can naturally leads to $\theta_{23}\simeq\frac{\pi}{4}$, which also appears as a remarkable feature of $P_{12}$ pattern (Fig.\ref{p12no}).

\section{Conclusion}
In this paper, we have performed an systematic investigation on the neutrino mass matrix $M_{\nu}$ with and one vanishing neutrino mass ($m_\text{lightest}=0$) and one texture equality($M_{\nu ab}=M_{\nu cd}$). Although the zero neutrino mass are strictly set in numerical analysis, our results is also well suited for the scenarios where an ultralight neutrino mass is assumed in the consideration of both experimental and theoretical sides. Using the latest neutrino oscillation and cosmological data, a phenomenological analysis are systematically proposed for all possible patterns.  It is found that twelve out of fifteen possible patterns are compatible with the experimental data at $3\sigma$ confidential level. In Fig. 1-15 and Table.3, we show the numerical results of viable patterns for normal order and inverted order of neutrino masses. 
For each viable pattern, allowed regions of  neutrino oscillation parameters $(\theta_{12},\theta_{23},\theta_{13},\delta)$, effective Majorana mass $m_{ee}$, and Jarlskog quantity $J_{CP}$ are presented.  We have summarized the main results in the numerical part. These interesting predictions are promising to be explored in the upcoming long-baseline neutrino oscillation experiment, neutrinoless double beta decay experiments and further cosmological study to the sum of neutrino masses. 

Finally, we discussed the flavor symmetry realization of texture structures, where  a concrete example has been illustrated in Ref.\cite{Lashin:2013xha} based on $S_{\mu-\tau}\times Z_{8}$ symmetry.  Inspired by this, we construct the model in the framework of Froggatt-Nielsen (FN) mechanism and subsequently realize the $P12$ pattern that naturally includes an ultralight neutrino mass.  The theoretical realization of other patterns deserves further study. Anyway, we except that a cooperation between theoretical study from the flavor symmetry viewpoint and a phenomenology study will help us reveal the underlying structure of massive neutrinos.

\section{Acknowledgments}
The work of Weijian Wang is supported by National Natural Science
Foundation of China under Grant Numbers 11505062, Special Fund of
Theoretical Physics under Grant Numbers 11447117 and Fundamental
Research Funds for the Central Universities.

\end{document}